\documentclass[
aps,
prd,
10pt,
showpacs,
amsmath,
amssymb,
twocolumn,
nofootinbib,
nobibnotes,
preprintnumbers
]{revtex4-1}
\usepackage{float}
\usepackage{mathrsfs,amsfonts}
\usepackage{mathtools}
\usepackage{bm}
\usepackage{tensor}
\usepackage{nicefrac}
\usepackage{url}
\usepackage{xcolor}
\usepackage[hyperindex,breaklinks,pdfusetitle]{hyperref}
\let\Tr\undefined

\DeclareMathOperator{\tr}{tr}
\DeclareMathOperator{\Tr}{Tr}
\DeclareMathOperator{\Det}{Det}

\begin{document}
\allowdisplaybreaks[1]
\title{Renormalization of generalized vector field models in curved spacetime}
\author{Michael S. Ruf}
\author{Christian F. Steinwachs}
\email{christian.steinwachs@physik.uni-freiburg.de}

\affiliation{Physikalisches Institut, Albert-Ludwigs-Universit\"at Freiburg,\\
Hermann-Herder-Stra\ss e~3, 79104 Freiburg, Germany}
%

\begin{abstract}
We calculate the one-loop divergences for different vector field models in curved spacetime. We introduce a classification scheme based on their degeneracy structure, which encompasses the well-known models of the nondegenerate vector field, the Abelian gauge field and the Proca field.
The renormalization of the generalized Proca model, which has important applications in cosmology, is more complicated. By extending standard heat-kernel techniques, we derive a closed form expression for the one-loop divergences of the generalized Proca model.
\end{abstract}


\maketitle


\section{Introduction}\label{SecIntro}

Most models of inflation and dynamical dark energy are based on scalar-tensor theories and $f(R)$ gravity, which have an additional propagating scalar degree of freedom. The one-loop quantum corrections to these models on an arbitrary background manifold have been derived for a general scalar-tensor theory in \cite{Shapiro1995, Steinwachs2011} and recently for $f(R)$ gravity in \cite{Ruf2018}. 

Aside from models based on an additional scalar field, vector fields have been studied in cosmology \cite{Novello1979,Davies1985,Ford1989, Kanno2008,Golovnev2008,Esposito-Farese2010,Heisenberg2014, Belokogne2016}. Most of these models are characterized by a nonminimal coupling of the vector field to gravity and are particular cases of the generalized Proca model.

The quantum corrections for the generalized Proca model are difficult to calculate and have been studied recently in \cite{Toms2015,Buchbinder2017} by different approaches. In this article, we use another approach, which allows us to derive the one-loop divergences for the generalized Proca model in a closed form.

We use a combination of the manifest covariant background field formalism and the heat kernel technique \cite{DeWitt1965, Atiyah1973, Gilkey1975, Ichinose1982, Abbott1982, Jack1984, Barvinsky1985, Avramidi2000, Vassilevich2003}.
This general approach can be applied to any type of field. The central object in this approach is the differential operator, which propagates the fluctuations of the fields. For most physical theories, this fluctuation operator acquires the form of a second order minimal (Laplace-type) operator. For this simple class of operators, a closed algorithm for the calculation of the one-loop divergences exists \cite{DeWitt1965}.
For nonminimal and higher order operators, a generalization of the Schwinger-DeWitt algorithm, which allows to reduce the calculation to the known case of the minimal second order operator, has been introduced in \cite{Barvinsky1985}. 
The direct application of the generalized Schwinger-DeWitt algorithm requires the nondegeneracy of the principal part---the highest-derivative term of the fluctuation operator. 

However, there are important cases, where the fluctuation operator has a degenerate principal part---notably $f(R)$ gravity \cite{Ruf2018} and the generalized Proca model considered in this article.
Therefore, we make use of the St\"uckelberg formalism \cite{Stueckelberg1938} to reformulate the generalized Proca model as a gauge theory such that standard heat-kernel techniques become applicable again. The price to pay is the introduction of a second metric tensor.  

This article is organized as follows:
In Sec. \ref{Sec:VectorModels}, we discuss different vector field models in curved spacetime. In particular, we introduce a classification based on their degeneracy structure.
In Sec. \ref{Sec:NonDegenerateVector}, we calculate the one-loop divergences for the nondegenerate vector field with an arbitrary potential.
In Sec. \ref{Sec:AbelianGaugeField}, we consider the case of the Abelian gauge field and calculate the one-loop divergences.
In Sec. \ref{Sec:ProcaModel}, we derive the one-loop divergences for the Proca model of the massive vector field. 
In Sec. \ref{Sec:GenVecModel}, we introduce the generalized Proca model, calculate the one-loop divergences in a closed form and present our main result.
In Sec. \ref{Sec:ChecksAndAppl}, we perform several reductions of our general result for the generalized Proca model to specific cases. These reductions provide strong cross checks of our general result and entail applications to cosmological models.
In Sec. \ref{Sec:ComparisonWithLit}, we compare our result and our method to previous calculations of the one-loop divergences for the generalized Proca model.  
Finally, in Sec. \ref{Sec:Conclusion}, we summarize our main results and give a brief outlook on their implications.

Technical details are provided in several appendices.
In Appendix \ref{App:GenForm}, we introduce the general formalism  and provide a collection of tabulated coincidence limits, which arise in the calculation.
In Appendix \ref{App:NDVT}, we present the details of the calculation for the one-loop divergences of the nondegenerate vector field considered in Sec. \ref{Sec:NonDegenerateVector}.
In Appendix \ref{App:TraceCalc}, we provide a detailed calculation of the most complex functional traces, which contribute to the one-loop divergences of the generalized Proca model considered in Sec. \ref{Sec:GenVecModel}.  
In Appendix \ref{App:Integrals}, we collect several important integral identities.
\vspace{9mm}
\begin{widetext}
	\begin{center}
	\begin{figure}[H]
		\centering
		\includegraphics{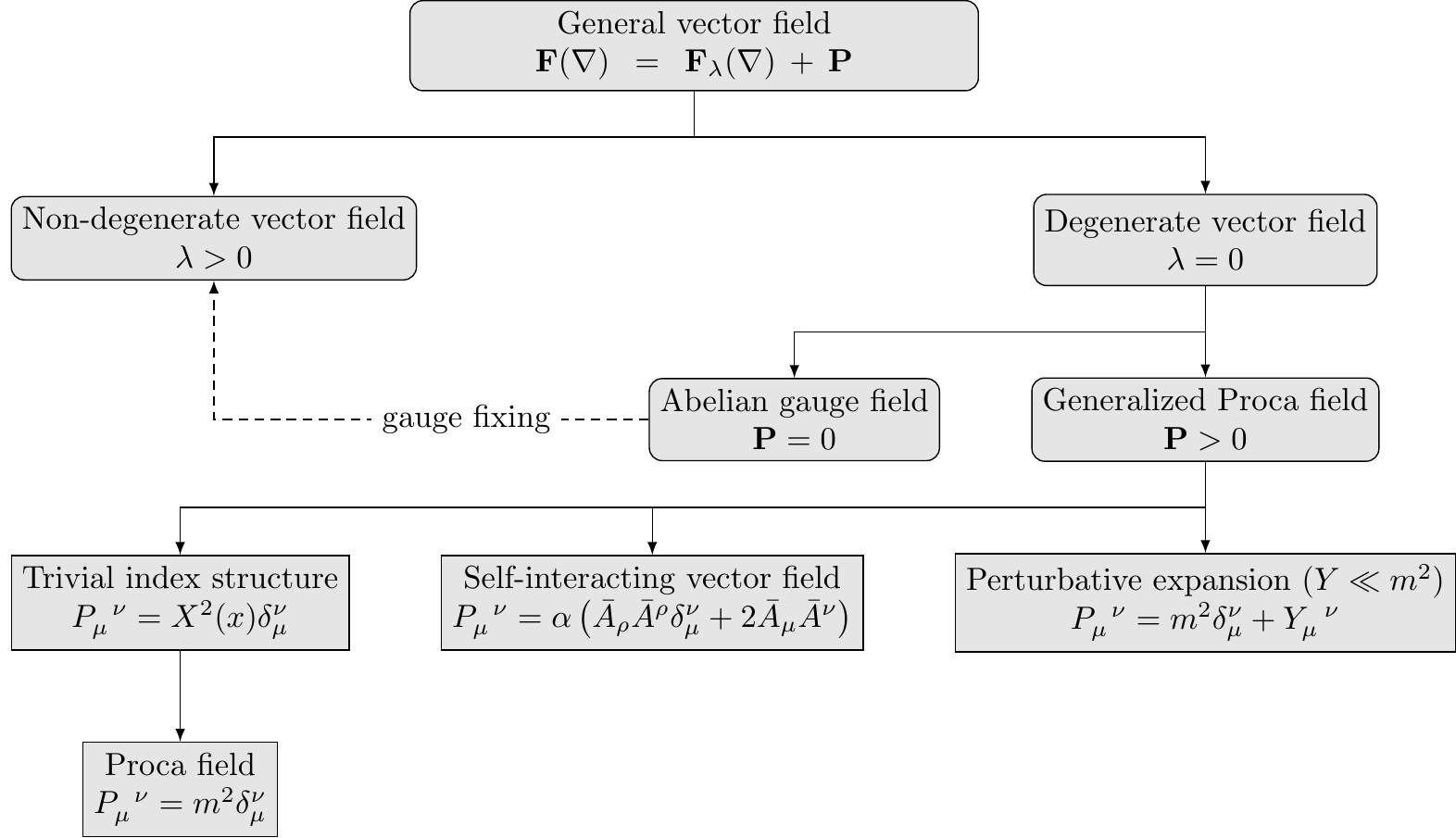}
		\caption{Overview of the different degeneracy classes and reductions of the generalized Proca model.}
		\label{Fig1}
	\end{figure}
	\end{center}
\end{widetext}

\section{Vector field models in curved spacetime: general structure}\label{Sec:VectorModels}
In this article we calculate the one-loop divergences for generalized vector field models in a curved spacetime.
The divergent part of the one-loop contribution to the effective action is defined by
\begin{align}
\Gamma_{1}^{\mathrm{div}}=\frac{1}{2}\left.\Tr\ln\mathbf{F}(\nabla)\right|^{\mathrm{div}}\,,\label{GamEffDiv1L}
\end{align}
where $\mathbf{F}(\nabla)$ is the differential operator that controls the propagation of the fluctuations. 
For the vector field models discussed in this work, this fluctuation operator has the particular form
\begin{align}
\mathbf{F}(\nabla)=\mathbf{F_{\lambda}}(\nabla)+\mathbf{P}\,,\label{OpVecGen}
\end{align}
where $\mathbf{P}$ is a potential with components $\tensor{P}{_{\mu}^{\nu}}$ and $\mathbf{F}_{\lambda}(\nabla)$ is a differential operator
 with components 
\begin{align}
	\tensor{\left[F_{\lambda}\right]}{_\mu^\nu}
	=\tensor{\left[\Delta_{\mathrm{H}}\right]}{_{\mu}^{\nu}}+\left(1-\lambda\right)\nabla_{\mu}\nabla^{\nu}\,.\label{OpFLambda}
\end{align}
Here, the Hodge operator on vector fields $\mathbf{\Delta}_{\mathrm{H}}$ is defined in terms of the positive definite Laplacian $\Delta$ by
\begin{align}
\tensor{\left[\Delta_{\mathrm{H}}\right]}{_{\mu}^{\nu}}\coloneqq \Delta\tensor{\delta}{_{\mu}^{\nu}}+\tensor{R}{_{\mu}^{\nu}},\qquad \Delta\coloneqq -g^{\mu\nu}\nabla_{\mu}\nabla_{\nu}\,.\label{Hodge}
\end{align}
In the terminology of \cite{Barvinsky1985}, the operator \eqref{OpVecGen} is called nonminimal for $\lambda\neq1$ and minimal for $\lambda=1$.
Note that if not indicated otherwise, derivative operators act on everything to their right.
The Hodge operator satisfies
\begin{align}
\tensor{\left[\Delta_{\mathrm{H}}\right]}{_{\mu}^{\nu}}\nabla_{\nu}=\nabla_{\mu}\Delta\,.
\end{align}
An important property of the general second order vector field operator \eqref{OpVecGen} is its degeneracy structure, which is controlled by the parameter $\lambda$ and the potential $\mathbf{P}$. The conditions $\lambda\geq0$ and $\mathbf{P}\geq0$ ensure that the operator $\mathbf{F}$ is positive semidefinite. Therefore, there are three different degeneracy classes:
\begin{enumerate}
	\item The nondegenerate vector field: $\lambda>0$, $\mathbf{P}\geq0$
	\item The Abelian gauge field: $\lambda=0$, $\mathbf{P}=0$
	\item The (generalized) Proca field: $\lambda=0$, $\mathbf{P}>0$
\end{enumerate}
The relation among the classes is depicted graphically in Fig. \ref{Fig1}.
Physically, the different classes correspond to inequivalent theories with a different number of propagating degrees of freedom. Mathematically, this is reflected by the degeneracy structure of the operator \eqref{OpVecGen}, which is discussed in the following sections in detail. In particular, there is no smooth transition between the classes in the limits $\lambda\to0$ and $\mathbf{P}\to0$. Therefore, the different classes have to be treated separately. 

\subsection{Degeneracy of the principal symbol}
An important structure in the theory of differential operators $\mathbf{F}(\nabla)$ is the leading derivative term---the principal part $\mathbf{D}(\nabla)$. Separating the principal part from the lower derivative terms $\mathbf{\Pi}(\nabla)$, the operator $\mathbf{F}(\nabla)$ takes the form
\begin{align}
\mathbf{F}(\nabla)=\mathbf{D}(\nabla)+\mathbf{\Pi}(\nabla).
\end{align}
Physically, the principal part $\mathbf{D}(\nabla)$ contains the information about the dominant ultraviolet behavior of the underlying theory. 
Therefore, it is the natural starting point for the generalized Schwinger-DeWitt method \cite{DeWitt1965, Barvinsky1985}, which relies on the expansion $\mathbf{D}\gg\mathbf{\Pi}$ of the associated propagator, schematically
\begin{align}
\frac{\mathbf{1}}{F}=\frac{\mathbf{1}}{D+\Pi}=\frac{\mathbf{1}}{D}-\frac{\mathbf{1}}{D}\mathbf{\Pi}\frac{\mathbf{1}}{D}+\cdots,\label{UVExp}
\end{align} 
where $\mathbf{1}/F$ denotes the inverse of the linear operator $\mathbf{F}$.
Essential for this perturbative treatment is the notion of background dimension $\mathfrak{M}$, which is understood as the mass dimension of the background tensorial coefficients of the differential operator. We write $\mathbf{F}={\cal O}(\mathfrak{M}^k)$ for any operator $\mathbf{F}$, which has at least background dimension $\mathfrak{M}^k$.
The expansion \eqref{UVExp} critically relies on the invertibility of $\mathbf{D}$, which can be discussed at the level of the principal symbol $\mathbf{D}(n)$, formally obtained by replacing derivatives $\nabla_{\mu}$ by a constant vector field $\mathrm{i} n_{\mu}$.
For the vector field operator \eqref{OpVecGen} the components of the  principal symbol read
\begin{align}	\tensor{D}{_\mu^\nu}(n)=n^2\left[\delta_{\mu}^{\nu}-\left(1-\lambda\right)\frac{\tensor{n}{_\mu}\tensor{n}{^\nu}}{n^2}\right],\label{eq:PPartVOP}
\end{align}
with $n^2\coloneqq n_{\rho}n^{\rho}$.   
The parameter $\lambda$ controls the degeneracy of the principal symbol, as can be seen easily from the determinant
\begin{align}
\det\mathbf{D}(n)=\lambda\left(n^2\right)^4\,.\label{DetPS}
\end{align}
For $\lambda=0$, the determinant vanishes and therefore $\mathbf{D}(n)$ is not invertible. 
The origin of this degeneracy can be traced back to the fact that for $\lambda=0$, the principal symbol has the structure of a projector on transversal vector fields. 
This motivates the distinction between the two classes $\lambda>0$ and $\lambda=0$.
For $\lambda=0$, the further distinction between the cases $\mathbf{P}=0$ and $\mathbf{P}>0$ is connected with a degeneracy at the level of the full operator $\mathbf{F}$, discussed in the next subsection.

\subsection{Gauge degeneracy}\label{SubSec:GaugeDegeneracy}
The degeneracy at the level of the full operator $\mathbf{F}$ is a general feature of any gauge theory. In the context of the vector field operator \eqref{OpVecGen}, the relevant gauge theory is defined by the Euclidean action for the Abelian gauge field $\tensor{A}{_{\mu}}(x)$,
\begin{align}
S[A]=\frac{1}{4}\int\mathrm{d}^4xg^{1/2} \mathcal{F}_{\mu\nu}\mathcal{F}^{\mu\nu}.\label{AbelianGaugeAction}
\end{align}
The Abelian field strength tensor $\mathcal{F}_{\mu\nu}$ is defined as
\begin{align}
\mathcal{F}_{\mu\nu}\coloneqq \nabla_{\mu}A_{\nu}-\nabla_{\nu}A_{\mu}\,.\label{AbelianFieldStrength}
\end{align}
The action \eqref{AbelianGaugeAction} is invariant under infinitesimal gauge transformations 
\begin{align}
\delta_{\varepsilon}A_{\mu}=\partial_{\mu}\varepsilon\,,
\end{align}
where $\varepsilon(x)$ is the infinitesimal local gauge parameter. Gauge invariance of \eqref{AbelianGaugeAction} implies the Noether identity
\begin{align}
\partial_{\mu}\frac{\delta S[A]}{\delta A_{\mu}(x)}=\nabla_{\mu}\left(g^{-1/2}\frac{\delta S[A]}{\delta A_{\mu}(x)}\right)=0\,.\label{Noether}
\end{align}
The components of the fluctuation operator $\mathbf{F}$ are obtained from the Hessian
\begin{align}
\tensor{F}{_\mu^\nu}(\nabla^x)\delta(x,x')\coloneqq g^{-1/2} g_{\mu\rho}\frac{\delta^2 S[A]}{\delta A_{\rho}(x)\delta A_{\nu}(x')}\,,
\end{align}
where the delta function is defined with zero density weight at $x$ and unit density weight at $x'$.
The fluctuation operator for the Abelian gauge field is given by $\mathbf{F}=\mathbf{F}_0$, which corresponds to the vector field operator \eqref{OpVecGen} with $\lambda=0$ and $\mathbf{P}=0$. The explicit components read
\begin{align}
\tensor{F}{_\mu^\nu}
=\tensor{\left[\Delta_{\mathrm{H}}\right]}{_{\mu}^{\nu}}+\nabla_{\mu}\nabla^{\nu}\,.\label{OpVecAbel}
\end{align}
Taking the functional derivative of \eqref{Noether} with respect to $A_{\nu}(x')$ yields the operator equation
\begin{align}
\nabla^{\mu}\tensor{F}{_\mu^\nu}=0\,.
\end{align}
This implies that for $\mathbf{P}=0$ the total fluctuation operator \eqref{OpVecGen} is degenerate -- not only its principle symbol.
Therefore, in case of a gauge degeneracy, in addition to the breakdown of the perturbative expansion \eqref{UVExp} associated to the degeneracy of $\mathbf{D}$, the inverse operator $\mathbf{1}/F$ does not even exist.  
In order to remove the gauge degeneracy we choose a gauge condition linear in $A_{\mu}$,
\begin{align}
\chi(A)=0\,.
\end{align}
The total gauge-fixed action $S_\mathrm{tot}\coloneqq S+S_{\mathrm{gb}}$ is obtained by adding the gauge breaking action 
\begin{align}
S_{\mathrm{gb}}=\frac{1}{2}\int\mathrm{d}^4xg^{1/2} \chi^2\,.
\end{align}
This leads to a modification of the associated Hessian, such that the resulting gauge-fixed operator $\mathbf{F}_\mathrm{tot}$ is nondegenerate 
\begin{align}
\tensor{\left[F_\mathrm{tot}\right]}{_\mu^\nu}(\nabla^x)\delta(x,x')=g^{-1/2} g_{\mu\rho}\frac{\delta^2S_{\mathrm{tot}}[A]}{\delta A_{\rho}(x)\delta A_{\nu}(x')}\,.\label{Ftot}
\end{align}
The inclusion of the gauge breaking action must be compensated by the corresponding ghost action
\begin{align}
S_{\mathrm{gh}}[\omega,\omega^{*}]=\int\mathrm{d}^4xg^{1/2}\omega^{*} Q\,\omega\,,
\end{align}
where $\omega^{*}(x)$ and $\omega(x)$ are anticommuting scalar ghost fields.
The ghost operator is defined as
\begin{align}
Q(\nabla^x)\delta(x,x')\coloneqq\frac{\delta}{\delta\varepsilon(x^\prime)}\chi\left(A(x)+\delta_{\varepsilon}A(x)\right)\,.\label{Qghost}
\end{align}
The divergent part of the one-loop contributions to the effective action is given by
\begin{align}
\Gamma_{1}^{\mathrm{div}}=\frac{1}{2}\left.\Tr_1\ln\mathbf{F}_{\mathrm{tot}}\right|^{\mathrm{div}}-\left.\Tr_0\ln \mathbf{Q}\right|^{\mathrm{div}}\,.\label{DivEffActGauge}
\end{align}

\section{The nondegenerate vector field}\label{Sec:NonDegenerateVector}
We first consider the vector field operator \eqref{OpVecGen} with a nondegenerate principal symbol and arbitrary potential
\begin{align}
\tensor{F}{_\mu^\nu}
=\tensor{\left[\Delta_{\mathrm{H}}\right]}{_{\mu}^{\nu}}+\left(1-\lambda\right)\nabla_{\mu}\nabla^{\nu}+\tensor{P}{_\mu^\nu},\qquad\lambda>0\,,\label{OpVecND}
\end{align}
Since for $\lambda\neq0$ the principal symbol of \eqref{OpVecND} is invertible, the generalized Schwinger-DeWitt algorithm can be used directly \cite{Barvinsky1985}. The power of this algorithm lies in its generality, as it is applicable to any type of field. However, instead of using the general algorithm, here the calculation can be essentially simplified by directly making use of an operator identity for $\mathbf{F}_{\lambda}$,
\begin{align}
\frac{\delta_{\mu}^{\nu}}{F_\lambda}=\frac{\delta_{\mu}^{\nu}}{\Delta_{\mathrm{H}}}-\gamma\nabla_{\mu}\frac{1}{\Delta^2}\nabla^{\nu}\,,\label{NDVFOpID}
\end{align}
where we have defined $\gamma\coloneqq(1-\lambda)/\lambda$.
Since $\mathbf{P}=\mathcal{O}\left(\mathfrak{M}^2\right)$ and $\mathbf{1}/F_{\lambda}=\mathcal{O}\left(\mathfrak{M}^0\right)$, we can make efficient use of \eqref{NDVFOpID}. For the calculation of the one-loop divergences, it is sufficient to expand the logarithm up to $\mathcal{O}\left(\mathfrak{M}^4\right)$,
\begin{align}
\left.\Tr_{1}\ln\mathbf{F}\right|^{\mathrm{div}}={}&\left.\Tr_{1}\ln\left(\mathbf{F}_{\lambda}+\mathbf{P}\right)\right|^{\mathrm{div}}\nonumber\\
={}&\left.\Tr_{1}\ln\mathbf{F}_{\lambda}\right|^{\mathrm{div}}+
\left.\Tr_{1}\left(\mathbf{P}\frac{\mathbf{1}}{F_{\lambda}}\right)\right|^{\mathrm{div}}\nonumber\\
&-\frac{1}{2}\left.\Tr_{1}\left(\mathbf{P}\frac{\mathbf{1}}{F_{\lambda}}\mathbf{P}\frac{\mathbf{1}}{F_{\lambda}}\right)\right|^{\mathrm{div}}\,.\label{ExpTRNDVF}
\end{align}
Inserting the operator identity \eqref{NDVFOpID}, the divergent contributions of the individual terms in \eqref{ExpTRNDVF} can be reduced to the evaluation of universal functional traces. The details of this calculation are provided in Appendix \ref{App:NDVT}. In this way, we find for the one-loop divergences of the nondegenerate vector field 
\begin{align}
\Gamma_1^{\mathrm{div}}={}&\frac{1}{32\pi^2\varepsilon}\int\mathrm{d}^4xg^{1/2}\left[\frac{11}{180}\mathcal{G}-\frac{7}{30}R_{\mu\nu}R^{\mu\nu}+\frac{1}{20}R^2\right.\nonumber\\
&+\left.\left(\frac{1}{6}+\frac{\gamma}{12}\right)RP-\left(1+\frac{\gamma}{6}\right)R_{\mu\nu}P^{\mu\nu}\right.\nonumber\\
&-\left.\left(\frac{1}{2}+\frac{\gamma}{4}+\frac{\gamma^2}{24}\right)P_{\mu\nu}P^{\mu\nu}-\frac{\gamma^2}{48}P^2\right]\,.\label{eq:EffactNonDegVF}
\end{align}
Here, we have defined the Gauss-Bonnet term  
\begin{align}
\mathcal{G}\coloneqq\tensor{R}{_\mu_\nu_\rho_\sigma}\tensor{R}{^\mu^\nu^\rho^\sigma}-4\,\tensor{R}{_\mu_\nu}\tensor{R}{^\mu^\nu}+R^2\,.\label{GaussBonnet}
\end{align}
The result \eqref{eq:EffactNonDegVF} is in agreement with \cite{Fradkin1982, Barvinsky1985}.\footnote{Apart from an overall minus sign, the transition to Lorentzian signature corresponds to the replacement $\mathbf{P}\to -\mathbf{P}$.}
Note that for $\mathbf{P}=0$, \eqref{eq:EffactNonDegVF} is independent of the parameter $\gamma$.
This calculation, as well as the calculation via the generalized Schwinger-DeWitt algorithm in \cite{Barvinsky1985}, both critically rely on the nondegeneracy of the principal symbol \eqref{DetPS}.

\section{The Abelian gauge field}\label{Sec:AbelianGaugeField}

The fluctuation operator $\mathbf{F}$ for the Abelian gauge field theory \eqref{AbelianGaugeAction} is given by
\begin{align}
 \tensor{F}{_\mu^\nu}
	=\tensor{\left[\Delta_{\mathrm{H}}\right]}{_{\mu}^{\nu}}+\nabla_{\mu}\nabla^{\nu}\label{AbelianOP}\,.
\end{align}
In view of the general operator \eqref{OpVecGen}, this corresponds to the case $\lambda=0$ and $\mathbf{P}=0$. 
As discussed in Sec. \ref{SubSec:GaugeDegeneracy}, the operator \eqref{AbelianOP} is degenerate due to the gauge symmetry of the action \eqref{AbelianGaugeAction}.
We choose a relativistic gauge condition with arbitrary gauge parameter $\eta$ to break the gauge degeneracy of the operator \eqref{AbelianOP}, 
\begin{align}
\chi(A)=-\frac{1}{\sqrt{1+\eta}}\nabla^{\mu}A_{\mu}\,.
\end{align}
According to \eqref{Ftot} and \eqref{Qghost}, the components of the gauge-fixed fluctuation operator $\mathbf{F}_{\mathrm{tot}}$ and the corresponding ghost operator $\mathbf{Q}$ read 
\begin{align}
\tensor{\left[F_{\mathrm{tot}}\right]}{_{\mu}^{\nu}}={}&\tensor{\left[\Delta_{\mathrm{H}}\right]}{_{\mu}^{\nu}}+\frac{\eta}{1+\eta}\nabla_{\mu}\nabla^{\nu},\label{GFOpAbGT}\\
 Q={}&\frac{1}{\sqrt{1+\eta}}\Delta\,.
\end{align}
Thus, the gauge-fixed fluctuation operator  \eqref{GFOpAbGT} falls into the class of nondegenerate vector fields \eqref{OpVecND} with $\mathbf{P}=0$ and $\lambda=1/(1+\eta)$. Therefore, the divergent part of the one-loop effective action can be calculated with the methods presented in Sec. \ref{Sec:NonDegenerateVector},
\begin{align}
 \Gamma^{\mathrm{div}}_{1}=\frac{1}{2}\left.\Tr_{1}\ln \mathbf{F}_{\mathrm{tot}}\right|^{\mathrm{div}}-\left.\Tr_0\ln \mathbf{Q}\right|^{\mathrm{div}}\,.\label{AbelianTraces}
\end{align}
 The first trace follows from \eqref{eq:EffactNonDegVF} for $\mathbf{P}=0$, while the ghost trace is evaluated directly with the standard result for a minimal second order operator \eqref{1LActionMinimal} and \eqref{SDWa2Coeff},
\begin{align}
\Gamma_1^{\mathrm{div}}=&\frac{1}{32\pi^2\varepsilon}\int{\rm d}^4x\,g^{1/2}\left( \frac{13}{180}\mathcal{G}-\frac{1}{5}\tensor{R}{_\mu_\nu}\tensor{R}{^\mu^\nu}+\frac{1}{15}\,R^2\right)\,.
\label{eq:EffactMaxwell}
\end{align}
The result \eqref{eq:EffactMaxwell} is in agreement with \cite{Barvinsky1985}.
Since the action for the Abelian vector field corresponds to a free theory, there are no contributions to the renormalization of the square of the field strength tensor \eqref{AbelianFieldStrength}. Note also that the result \eqref{eq:EffactMaxwell} is independent of the gauge parameter $\eta$.

\section{The Proca field}\label{Sec:ProcaModel}
The Proca action for the massive vector field in curved spacetime is given by the action of the Abelian gauge field \eqref{AbelianGaugeAction} supplemented by a mass term \cite{Proca1936},
\begin{align}
S[A]=\int\mathop{}\!\mathrm{d}^4x\,g^{1/2}\left(\frac{1}{4}\,\mathcal{F}_{\mu\nu}\mathcal{F}^{\mu\nu}+\frac{1}{2}m^2A_{\mu}A^{\mu}\right)\,.\label{ActProc}
\end{align}
The mass term breaks the gauge symmetry. The Hessian of \eqref{ActProc} leads to the fluctuation operator
\begin{align}
\tensor{F}{_\mu^\nu}
	\coloneqq\tensor{\left[\Delta_{\mathrm{H}}\right]}{_{\mu}^{\nu}}+\nabla_{\mu}\nabla^{\nu}+m^2\delta_{\mu}^{\nu}\,,\label{ProcOP}
\end{align}
which corresponds to the case $\lambda=0$ and $\mathbf{P}=m^2\mathbf{1}$ of the general vector field operator \eqref{OpVecGen}, that is $\mathbf{F}=\mathbf{F}_0+m^2\mathbf{1}$.
The mass term in the Proca operator \eqref{ProcOP} breaks the gauge degeneracy of the gauge field operator $\mathbf{F}_0$.
Nevertheless, the principal part of the Proca operator \eqref{ProcOP} is still degenerate. This degeneracy cannot be removed by a gauge fixing---in contrast to the Abelian gauge field. 
Similar to \eqref{NDVFOpID}, there is an operator identity for the Proca field
\begin{align}
\frac{\delta_{\mu}^{\nu}}{F_0+m^2}=\left(\tensor*{\delta}{_\mu^\rho}-\frac{\nabla_{\mu}\nabla^{\rho}}{m^2}\right)\frac{\delta_{\rho}^{\nu}}{\Delta_{\mathrm{H}}+m^2}\,.\label{eq:MassivWard}
\end{align}
Taking the trace of the logarithm on both sides of \eqref{eq:MassivWard}
and using that the divergent part of the vector trace can be converted into a contribution from a scalar trace
\begin{align}
\Tr_1\ln\left.\left(\delta_{\mu}^{\nu}-\frac{\nabla_{\mu}\nabla^{\nu}}{m^2}\right)\right|^{\mathrm{div}}=\left.\Tr_0\ln\left(\Delta+m^2\right)\right|^{\mathrm{div}}\,,\label{eq:VecToScal}
\end{align}
the divergent part of the one-loop effective action is reduced to the vector and scalar traces of two minimal second order operators,
\begin{align}
		\Gamma_{1}^{\mathrm{div}}={}&\frac{1}{2}\left.\Tr_1\ln\left(\mathbf{F}_0+m^2\mathbf{1}\right)\right|^{\mathrm{div}}\nonumber\\
		 ={}&\frac{1}{2}\left.\Tr_1\ln\left(\mathbf{\Delta}_{\mathrm{H}}+m^2\mathbf{1}\right)\right|^{\mathrm{div}}\nonumber\\
		 &-\frac{1}{2}\left.\Tr_0\ln\left(\Delta+m^2\right)\right|^{\mathrm{div}}\,.\label{eq:ProcaEffactDec} 	
\end{align}
The vector and scalar traces in \eqref{eq:ProcaEffactDec} can be calculated directly with the closed form algorithm \eqref{1LActionMinimal}--\eqref{SDWa2Coeff}. The final result for the one-loop divergences of the Proca model \eqref{ActProc} reads
\begin{align}
		\Gamma_1^{\mathrm{div}}=
		&\frac{1}{32\pi^2\varepsilon}\int\mathrm{d}^4 x\,g^{1/2}\left( \frac{1}{15}\mathcal{G}-\frac{13}{60}\tensor{R}{_\mu_\nu}\tensor{R}{^\mu^\nu}\right.\nonumber\\
		&\left.+\frac{7}{120}R^2-\frac{1}{2}m^2 R-\frac{3}{2}m^4\right)\,.\label{eq:EffactProca}
	\end{align}
The result \eqref{eq:EffactProca} is in agreement with \cite{Barvinsky1985, Buchbinder2012}. It has a clear physical interpretation. The effective action of the massive vector field in four dimensions is that of a four component vector field minus one scalar mode, since the Proca field has $4-1=3$ propagating degrees of freedom. Therefore, it is clear that \eqref{eq:EffactProca} does not reproduce the result for the Abelian gauge field \eqref{eq:EffactMaxwell} in the limit $m\to0$, as the Abelian gauge field has only $4-2=2$ propagating degrees of freedom. At the level of the functional traces, this can formally be seen as follows: while the scalar operator for the longitudinal mode of the Proca field in \eqref{eq:ProcaEffactDec} indeed reduces to the ghost operator of the Abelian gauge field in \eqref{AbelianTraces} in the limit $m\to0$, the ghost trace in \eqref{AbelianTraces} is subtracted twice compared to the trace of the longitudinal mode in \eqref{eq:ProcaEffactDec}.

\section{The generalized Proca field}\label{Sec:GenVecModel}
The generalized Proca model is defined by the action
	\begin{align}
	S[ A]={}&\int\mathrm{d}^4x\,g^{1/2}\left(\frac{1}{4}\tensor{\mathcal{F}}{_\mu_\nu}\tensor{\mathcal{F}}{^\mu^\nu}+\frac{1}{2}\tensor{M}{^\mu^\nu}\tensor{ A}{_\mu}\tensor{ A}{_\nu}\right)\,.\label{ActGenProca}
	\end{align}
The action is that of the Proca field \eqref{ActProc}, but with the scalar mass term $m^2$ generalized to an arbitrary positive definite and symmetric  background mass tensor $M^{\mu\nu}$.
The background mass tensor $M^{\mu\nu}$ is completely general and might be constructed from external background fields. Of particular interest are curvature terms
\begin{align}
\tensor{M}{^\mu^\nu}=\zeta_1\tensor{R}{^\mu^\nu}+\zeta_2R\tensor{g}{^\mu^\nu}\,,
\end{align}
which arise in cosmological models \cite{Davies1985,Ford1989,Novello1979,Kanno2008,Golovnev2008,Esposito-Farese2010,Heisenberg2014, Belokogne2016}. 
The fluctuation operator for the generalized Proca theory \eqref{ActGenProca} reads
\begin{align}
\tensor{F}{_{\mu}^{\nu}}=\tensor{\left[\Delta_{\mathrm{H}}\right]}{_{\mu}^{\nu}}+\nabla_{\mu}\nabla^{\nu}+\tensor{M}{_{\mu}^{\nu}}\,.\label{OpGenProca}
\end{align}
The operator \eqref{OpGenProca} corresponds to the general operator \eqref{OpVecGen} for $\lambda=0$ and $\tensor{P}{_{\mu}^{\nu}}=g_{\mu\rho}M^{\rho\nu}>0$.
The standard techniques for the calculation of the one-loop divergences are not directly applicable to the degenerate operator of the generalized Proca theory \eqref{OpGenProca}. In particular, for the generalized Proca operator, there is no simple analogue of the operator identity \eqref{eq:MassivWard} for the Proca operator. Therefore, we adopt a different strategy and first reformulate the generalized Proca theory as a gauge theory by making use of the St\"uckelberg formalism. In this formulation, the generalized background mass tensor $M^{\mu\nu}$ plays a double role as potential in the vector sector and as metric in the scalar St\"uckelberg sector. For this effective ``bimetric'' formulation, the standard heat-kernel techniques are applicable and the one-loop divergences of the generalized Proca action \eqref{ActGenProca} are obtained in a closed form. 

\subsection{Weyl transformation and bimetric formalism}\label{SecConfTrafoStueckTrick} 
The calculations are simplified by performing a Weyl transformation of the background metric 
\begin{align}
\tensor{\hat{g}}{_\mu_\nu}\coloneqq\frac{1}{\mu^2}\left[\det\left(\tensor*{M}{^{\mu\nu}}\tensor{g}{_{\nu\rho}}\right)\right]^{\nicefrac{1}{4}}\tensor{g}{_\mu_\nu}\,.\label{ConfTrafoMetric}
\end{align}
Here, $\mu$ is an auxiliary mass parameter, introduced for dimensional reasons. Note that in what follows indices are raised and lowered only with the metric $\hat{g}_{\mu\nu}$.
Since the kinetic term is invariant under a Weyl transformation, we find
	\begin{align}
	S=\int\mathrm{d}^4x\,\hat{g}^{1/2}\left[\frac{1}{4}\tensor{\hat{g}}{^\mu^\alpha}\tensor{\hat{g}}{^\nu^\beta}\tensor{\mathcal{F}}{_\mu_\nu}\tensor{\mathcal{F}}{_\alpha_\beta}+\frac{\mu^2}{2}\tensor{\left(\tilde{g}^{\,-1}\right)}{^\mu^\nu}\tensor{ A}{_\mu}\tensor{ A}{_\nu}\right]\,.\label{ActGenProcConfMet}
	\end{align}
In the second term, we have defined 
	\begin{align}
\tensor{\left(\tilde{g}^{\,-1}\right)}{^\mu^\nu}\coloneqq \mu^2\left[\det\left(\tensor*{M}{^{\mu\nu}}\tensor{g}{_{\nu\rho}}\right)\right]^{-1/2}\tensor{M}{^\mu^\nu}\,,\label{DefTildeg}
\end{align}
which is the inverse of the new metric $\tensor{\tilde{g}}{_{\rho\nu}}$.
In this way, formally the dependency on the original general mass tensor $M^{\mu\nu}$ has been replaced by a standard mass term.
By construction, we have the important relations
\begin{align}
\det \tensor{\tilde{g}}{_{\mu\nu}}=\det \tensor{\hat{g}}{_{\mu\nu}}\,,\quad \hat{\nabla}_{\rho}\det\tensor{\tilde{g}}{_{\mu\nu}}=0\,.\label{DetRelMetTwid}
\end{align}
We define the Christoffel connection associated with $\tensor{\tilde{g}}{_{\mu\nu}}$,
\begin{align} \tensor{\tilde{\Gamma}}{^{\rho}_{\mu\nu}}=\frac{1}{2}\tensor{\left(\tilde{g}^{-1}\right)}{^{\rho\alpha}}\left(\tensor{\partial}{_\mu}\tensor{\tilde{g}}{_{\alpha\nu}}+\tensor{\partial}{_\nu}\tensor{\tilde{g}}{_{\mu\alpha}}-\tensor{\partial}{_\alpha}\tensor{\tilde{g}}{_{\mu\nu}}\right)\,.
\end{align}
A natural structure is the difference tensor
\begin{align}
\tensor{\delta\Gamma}{^\lambda_\mu_\nu}\coloneqq{}&\tensor{\tilde{\Gamma}}{^\lambda_\mu_\nu}-\tensor{\hat{\Gamma}}{^\lambda_\mu_\nu}\nonumber\\
={}&\frac{1}{2}\tensor{\left(\tilde{g}^{\,-1}\right)}{^\lambda^\alpha}\left(\tensor{\hat{\nabla}}{_\mu}\tensor{\tilde{g}}{_\nu_\alpha}+\tensor{\hat{\nabla}}{_\nu}\tensor{\tilde{g}}{_\mu_\alpha}-\tensor{\hat{\nabla}}{_\alpha}\tensor{\tilde{g}}{_\mu_\nu}\right)\,.\label{DelGam}
\end{align}
By construction, the difference tensor satisfies
\begin{align}
\tensor{\delta\Gamma}{^\alpha_\mu_\alpha}={}&\frac{1}{2}\tensor{\left(\tilde{g}^{-1}\right)}{^\alpha^\beta}\hat{\nabla}_{\mu}\tensor{\tilde{g}}{_{\alpha\beta}}\nonumber\\
={}&\left(\det\tensor{\tilde{g}}{_{\rho\sigma}}\right)^{-1/2}\hat{\nabla}_\mu\left(\det \tensor{\tilde{g}}{_{\rho\sigma}}\right)^{1/2}=0\,.\label{TrDiffTensor}
\end{align}
The Ricci curvatures of the new metric $\tensor{\tilde{g}}{_\mu_\nu}$ are given by 
\begin{align}
\tilde{R}={}&\tensor{\left(\tilde{g}^{\,-1}\right)}{^\mu^\nu}\tensor{\tilde{R}}{_\mu_\nu}\,,\\
\tensor{\tilde{R}}{_\mu_\nu}={}&\tensor{\hat{R}}{_\mu_\nu}
-\tensor{\delta\Gamma}{^\alpha_\beta_\mu}\tensor{\delta\Gamma}{^\beta_\alpha_\nu}+\tensor{\hat{\nabla}}{_\alpha}\tensor{\delta\Gamma}{^\alpha_\mu_\nu}\,.
\end{align}
In the following, when we work with the two metrics $\tensor{\hat{g}}{_{\mu\nu}}$ and $\tensor{\tilde{g}}{_{\mu\nu}}$, indices are raised and lowered exclusively with the metric $\hat{g}_{\mu\nu}$.

\subsection{St\"uckelberg formalism}
As we have discussed in the context of the Abelian gauge field, the gauge symmetry is responsible for the degeneracy of the total fluctuation operator. Therefore, a gauge fixing is required to remove this degeneracy. At the same time, the gauge fixing can be used to also remove the degeneracy of the principal symbol.
A similar mechanism works in the case of the generalized Proca model, when artificially rewritten as a gauge theory, which is realized by the St\"uckelberg formalism.
The St\"uckelberg scalar field $\varphi$ is introduced by the shift
\begin{align}
	\tensor{ A}{_\mu}\to \tensor{ A}{_\mu}+\frac{1}{\mu}\partial_\mu\varphi\,.\label{Stueckelberg}
\end{align}
In terms of the vector field $A_{\mu}$ and the St\"uckelberg scalar $\varphi$, the action \eqref{ActGenProcConfMet} is given by
\begin{align}
S[ A,\varphi]={}&\int\mathrm{d}^4x\,\hat{g}^{1/2}\left[\frac{1}{4}\tensor{\hat{g}}{^\mu^\alpha}\tensor{\hat{g}}{^\nu^\beta}\tensor{\mathcal{F}}{_\mu_\nu}\tensor{\mathcal{F}}{_\alpha_\beta}\right.\nonumber\\
&\left.+\frac{\mu^2}{2}\tensor{\left(\tilde{g}^{\,-1}\right)}{^\mu^\nu}\tensor{ A}{_\mu}\tensor{ A}{_\nu}+\mu\tensor{\left(\tilde{g}^{\,-1}\right)}{^\mu^\nu}\tensor{ A}{_\mu}\partial_\nu\varphi\right.\nonumber\\
&\left.+\frac{1}{2}\tensor{\left(\tilde{g}^{\,-1}\right)}{^\mu^\nu}\partial_\nu\varphi\partial_\nu\varphi\right]\,.\label{ActStueckTwoField}
\end{align}
This action has a gauge symmetry as it is invariant under the simultaneous infinitesimal gauge transformations
\begin{align}
\delta_{\varepsilon}\tensor{ A}{_\mu}=\partial_\mu\varepsilon,\qquad \delta_{\varepsilon}\varphi=-\mu\varepsilon\,.
\end{align}
We choose a one-parameter family of gauge conditions
\begin{align}
\chi[A,\varphi]=-\frac{1}{\sqrt{1+\eta}}\left(\hat{\nabla}_\mu\tensor{ A}{^\mu}+\eta \mu\varphi\right)\,.\label{GaugeCondition}
\end{align}
In particular, \eqref{GaugeCondition} interpolates between the original vector field theory  \eqref{ActGenProcConfMet} with $\varphi=0$ ($\eta\to\infty$) and the Lorentz gauge ($\eta=0$). The corresponding gauge breaking action reads
\begin{align}
	S_{\mathrm{gb}}[A,\varphi]=\frac{1}{2}\int\mathrm{d}^4x\,{\hat{g}}^{1/2}\chi^2\,.
\end{align}
The ghost operator is obtained from \eqref{GaugeCondition},
\begin{align}
Q(\hat{\nabla})={}&\hat{\Delta}+\eta \mu^2\,.\label{GhostOp}
\end{align}
For simplicity, we choose the  Lorentz gauge, $\eta=0$.\footnote{The calculation can be performed for the general $\eta$-family of gauges \eqref{GaugeCondition}. By using \eqref{NDVFOpID}, it can be seen already at the level of the functional traces that all $\eta$-dependent terms cancel and the one-loop divergences are independent of the gauge parameter $\eta$.} 
In terms of the generalized two-component field 
\begin{align}
\tensor{\phi}{^{A}}=\left[\begin{array}{c}A_{\mu}\\\varphi\end{array}\right]\,,
\end{align}
the gauge-fixed action acquires the block form
\begin{align}
S[A,\varphi]+S_{\mathrm{gb}}[A,\varphi]={}&\int\mathrm{d}^4x\,g^{1/2}\tensor{\phi}{^{A}}\tensor{F}{_{AB}}\tensor{\phi}{^{B}},
\end{align}
where the block matrix fluctuation operator $\mathbf{F}$ has components $\tensor{F}{^A_B}=\gamma^{AC}F_{CB}$. The components of the (dedensitized) inverse configuration space metric $\gamma^{AB}$ are given by
\begin{align}
\gamma^{AB}=\begin{bmatrix}
\hat{g}_{\mu\nu}&\\
& 1
\end{bmatrix}\,.
\end{align}
Splitting the fluctuation operator according to the number of derivatives, it can be represented as 
\begin{align}
\mathbf{F}=\mathbf{D}+\mathbf{\Pi}\,,\label{FBlockOp}
\end{align}
with the block matrix structure
\begin{align}
\mathbf{D}={}&\begin{bmatrix}
\mathbf{D}_1&\\
&\mathbf{D}_0
\end{bmatrix}\,,
\qquad \mathbf{\Pi}=\begin{bmatrix}
&\mathbf{\Pi}^\dagger\\
\mathbf{\Pi}&
\end{bmatrix}\,.\label{BlockOp}
\end{align}
The components of the operators in \eqref{BlockOp} are given by
\begin{align}
\tensor{\left[D_1\right]}{_\mu^\nu}\coloneqq{}&\tensor{\left[\hat{\Delta}_{\mathrm{H}}\right]}{_\mu^\nu}+\mu^2\tensor*{\left(\tilde{g}^{\,-1}\right)}{_\mu^\nu}\,,\nonumber\\
\tensor*{D}{_0}\coloneqq{}&-\hat{\nabla}_\mu\tensor{\left(\tilde{g}^{\,-1}\right)}{^\mu^\nu}\hat{\nabla}_\nu\,,\nonumber\\
\tensor*{\Pi}{^\nu}\coloneqq{}&-\mu\hat{\nabla}_\mu\tensor{\left(\tilde{g}^{\,-1}\right)}{^\mu^\nu}\,,\label{VOpComp}
\end{align}
where $\tensor*{\Pi}{^\dagger_\mu}=\mu\tensor*{\left(\tilde{g}^{\,-1}\right)}{_\mu^\nu}\hat{\nabla}_\nu\,,$ denotes the formal adjoint of $\tensor*{\Pi}{^\nu}$ with respect to the inner product on the space of vectors. The component $D_0$ can be simplified by using \eqref{TrDiffTensor} and defines the scalar Laplace operator with respect to the metric $\tensor{\tilde{g}}{_{\mu\nu}}$,
\begin{align}
D_0=-\tensor{\left(\tilde{g}^{\,-1}\right)}{^\mu^\nu}\tilde{\nabla}_\mu\tilde{\nabla}_\nu\eqqcolon\tilde{\Delta}\,.
\end{align}
Let us briefly discuss what we have achieved by the St\"uckelberg formalism. 
The $4-1=3$ propagating degrees of freedom of the original generalized Proca field have been converted into the $4+1-2=3$ propagating degrees of freedom, corresponding to those of a vector field, a scalar, and two scalar ghosts fields.  
In contrast to the principal part of the original generalized Proca operator \eqref{OpGenProca}, the additional gauge freedom present in the St\"uckelberg formalism has been used to render the principal part $\mathbf{D}$ of the scalar-vector block operator \eqref{FBlockOp} nondegenerate and, in particular, minimal---the price to pay is the introduction of the second metric $\tilde{g}_{\mu\nu}$.

\subsection{One-loop effective action}
In order to calculate the one-loop effective action
\begin{align}
\Gamma_{1}=\frac{1}{2}\Tr\ln \mathbf{F}-	\Tr_0 \ln \mathbf{Q}\,,
\end{align}
we expand $\mathbf{F}$ around $\mathbf{D}$. Perturbation theory in $\mathbf{\Pi}$ is efficient as $\mathbf{\Pi}={\cal O}(\mathfrak{M}^1)$. 
Expanding $\Tr\ln \mathbf{F}$ up to ${\cal O}(\mathfrak{M}^4)$, we obtain
\begin{align}
\Tr\ln \mathbf{F}={}&\Tr_1\ln\mathbf{D}_{1}+\Tr_0 \ln \mathbf{D}_0-T_2-\frac{1}{2}T_4\,,\label{TrLogExp}
\end{align}
where $T_i=\mathcal{O}\left(\mathfrak{M}^i\right)$, $i=2,4$ denote the following traces
\begin{align}
T_2\coloneqq{}&\frac{1}{2}\Tr\left[\left(\mathbf{\Pi}\frac{\mathbf{1}}{D}\right)^2\right]=\Tr_0 \left(\tensor{\Pi}{^\alpha}\frac{\tensor*{\delta}{_\alpha^\beta}}{D_1}\tensor*{\Pi}{^\dagger_\beta}\frac{1}{D_0}\right)\label{T1Trace}\,,\\
T_4\coloneqq{}& \frac{1}{2}\Tr\left[\left(\mathbf{\Pi}\frac{\mathbf{1}}{D}\right)^4\right]=\Tr_0\left[\left( \tensor{\Pi}{^\alpha}\frac{\tensor*{\delta}{_\alpha^\beta}}{D_1}\tensor*{\Pi}{^\dagger_\beta}\frac{1}{D_0}\right)^2\right].\label{T2Trace}
\end{align}
Note that odd powers in the expansion are zero, because the block matrix $\mathbf{\Pi}\,\mathbf{1}/D$ has only zeros on the diagonal and that we have used the cyclicity of the trace to convert vector traces into scalar traces, for example:
\begin{align}
\Tr_1\left(\tensor*{\Pi}{^\dagger_\mu}\frac{1}{D_0} \tensor{\Pi}{^\alpha}\frac{\tensor*{\delta}{_\alpha^\nu}}{D_1}\right)=\Tr_0\left( \tensor{\Pi}{^\alpha}\frac{\tensor*{\delta}{_\alpha^\beta}}{D_1}\tensor*{\Pi}{^\dagger_\beta}\frac{1}{D_0}\right)\,.
\end{align}
The first two traces in \eqref{TrLogExp}, together with the contribution of the ghost operator \eqref{GhostOp}, are evaluated directly with \eqref{1LActionMinimal}--\eqref{SDWa2Coeff}. Their sum reads
\begin{align}
&\Tr_1\ln \mathbf{D}_{1}+\Tr_0 \ln \mathbf{D}_0-2\Tr_0\ln \mathbf{Q}\nonumber\\
={}&\frac{1}{16\pi^2\varepsilon}\int\mathrm{d}^4x\hat{g}^{1/2}\left[\frac{1}{15}\mathcal{\hat{G}}-\frac{1}{60}\left(\tilde{g}^{-1}\right)^{\mu\nu}\left(\tilde{g}^{-1}\right)^{\rho\sigma}\tilde{R}_{\mu\rho}\tilde{R}_{\nu\sigma}\right.\nonumber\\
&\left.-\frac{1}{120}\tilde{R}^2-\frac{1}{2}\mu^4\tr\left(\tilde{g}^{-2}\right)-\mu^2\left(\tilde{g}^{-1}\right)^{\mu\nu}\hat{R}_{\mu\nu}\right.\nonumber\\
&\left.+\frac{1}{6}\mu^2\tr\left(\tilde{g}^{-1}\right)\hat{R}-\frac{1}{5}\hat{R}_{\mu\nu}\hat{R}^{\mu\nu}+\frac{1}{15}\hat{R}^2\right]\,,\label{GenProcMinContributions}
\end{align}
where we have introduced the abbreviation
\begin{align}
	\left(\tilde{g}^{-n}\right)^{\mu\nu}\coloneqq\left(\tilde{g}^{-1}\right)^{\mu\alpha_1}\tensor{\left(\tilde{g}^{-1}\right)}{_{\alpha_1}^{\alpha_2}}\cdots\tensor{\left(\tilde{g}^{-1}\right)}{_{\alpha_{n-1}}^{\nu}}\,.
\end{align}
In \eqref{GenProcMinContributions}, we used that the Euler characteristic
\begin{align}
\chi(\mathcal{M})\coloneqq\frac{1}{32\pi^2}\int\mathrm{d}^4x\,\hat{g}^{1/2}\mathcal{\widehat{G}}=\frac{1}{32\pi^2}\int\mathrm{d}^4x\,\tilde{g}^{1/2}\mathcal{\tilde{G}}\,,
\end{align}
defined in terms of the Gauss-Bonnet term \eqref{GaussBonnet}, is a topological invariant and therefore independent of the metric. This allows to combine both contributions from $\hat{g}_{\mu\nu}$ and $\tilde{g}_{\mu\nu}$ in \eqref{GenProcMinContributions}.
The evaluation of the divergent contributions from the remaining traces \eqref{T1Trace} and \eqref{T2Trace} constitutes the most complex part of the calculation. Here, we only sketch the major steps. The details are provided in Appendix \ref{App:TraceCalc}.
There are two main complications associated with the evaluation of the divergent parts of the traces \eqref{T1Trace} and \eqref{T2Trace}. First, the traces \eqref{T1Trace} and \eqref{T2Trace} involve propagators $\delta_{\mu}^{\nu}/D_{1}$ and $1/D_0$ with different spin. Second, the propagators are defined with respect to different metrics. Therefore, we have to explicitly perform the convolution of the corresponding kernels
\begin{align}
T_2={}\int\mathrm{d}^{2\omega}x\,\mathrm{d}^{2\omega}x^{\prime}&\left[
\Pi^{\nu}\frac{\delta^{\mu^{\prime}}_\nu}{D_1}\hat{\delta}(x,x^{\prime})\right]\left[\vphantom{\frac{\delta^{\mu^{\prime}}_\nu}{D_1}}\Pi^{\dagger}_{\mu'}\frac{1}{D_0}\tilde{\delta}(x^{\prime},x)\right],\label{ProductTraces}
\end{align}
where we have defined $\omega\coloneqq d/2$.
Inserting the Schwinger-DeWitt representation for the kernels of $\mathbf{1}/D_1$ and $\mathbf{1}/D_0$, provided in Appendix \ref{App:TraceCalc}, the traces $T_{2}$ and $T_4$ are ultimately reduced to Gaussian integrals. In case of a single metric this procedure has been outlined in \cite{Barvinsky1985}. 
In the case of two metric structures, the problem becomes more complicated and has been discussed in \cite{Barvinsky1989}. The resulting Gaussian integral is
\begin{align}
T_2={}&\frac{1}{\left(4\pi\right)^{2\omega}}\int\mathrm{d}^{2\omega}x\,\hat{g}^{1/2}\int_{0}^{\infty}\mathrm{d}u\,u^{\omega-2}\int_{0}^{\infty}\frac{\mathrm{d}s}{s^{\omega-1}}\nonumber\\
&\times\int\left(\prod_{\mu=1}^{2\omega}\mathrm{d}\hat{\sigma}^{\mu}\right)\Psi(\hat{\sigma}^{\mu}) \exp\left(-\frac{1}{4}\tensor{G}{_\alpha_\beta}\tensor{\hat{\sigma}}{^\alpha}\tensor{\hat{\sigma}}{^\beta}\right)\,.\label{T2GaussGuRep}
\end{align}
Here, we have introduced the ``interpolation metric'' 
\begin{align}
\tensor{G}{_\mu_\nu}(u)\coloneqq{}&\tensor{\hat{g}}{_\mu_\nu}+u\tensor{\tilde{g}}{_\mu_\nu}\,.
\end{align}
The function $\Psi(u,s|\hat{\sigma}^{\mu})$ in the integrand of \eqref{T2GaussGuRep} is the result of the covariant Taylor expansion in $\hat{\sigma}^{\mu}$
\begin{align}
\Psi(u,s|\hat{\sigma}^{\mu})\coloneqq\hat{g}^{1/2}\sum_{k=0}^\infty \Psi^{(k)}_{(\mu_1\dots\mu_k)}(u,s)\hat{\sigma}^{\mu_1}\cdots\hat{\sigma}^{\mu_k}\,,
\end{align}
where $\hat{\sigma}^{\mu}(x,x^{\prime})$ is tangent to the geodesic connecting $x$ with $x^{\prime}$ at the point $x$.
Note that the background field dependent coefficients $\Psi^{(k)}_{(\mu_1...\mu_k)}(u,s)$ only involve positive powers of the parameters $u$ and $s$. In $d=2\omega=4$ dimensions, only terms of the integrand with total $s$-dependency $1/s$ contribute to the logarithmically divergent part.\footnote{The choice of the parametrization in terms of $s$ and $u$ guarantees that the divergent contribution is isolated in the $s$-integration, whereas the $u$-integration is finite.} Therefore, in view of \eqref{T2GaussGuRep}, the divergent contributions originate from the $s$-independent parts of $\Psi$. Finally, the Gaussian integrals in \eqref{T2GaussGuRep} are evaluated 
\begin{align}
&\frac{1}{\left(4\pi\right)^{2}}\int\left(\prod_{\mu=1}^{4}\mathrm{d}\hat{\sigma}^{\mu}\right)\,\hat{\sigma}^{\mu_1}\cdots\hat{\sigma}^{\mu_{2k}} \exp\left(-\frac{1}{4}\tensor{G}{_\alpha_\beta}\tensor{\hat{\sigma}}{^\alpha}\tensor{\hat{\sigma}}{^\beta}\right)\nonumber\\
={}&\frac{1}{G^{1/2}}\tensor*{\left[\operatorname{sym}_{k}\left(G^{-1}\right)\right]}{^{\mu_1\dots\mu_{2k}}}\,.\label{GaussInt}
\end{align}
Here, we have introduced 
the  $k$th totally symmetrized power of a general rank two tensor $T^{\mu\nu}$, defined by
\begin{align}
\tensor*{\left[\operatorname{sym}_{k}\left(T\right)\right]}{^{\mu_1\dots\mu_{2k}}}\coloneqq{}&\frac{(2k)!}{2^k k!}T^{(\mu_1\mu_2}\cdots T^{\mu_{2k-1}\mu_{2k})}\,.
\end{align}  
After evaluation of the Gaussians \eqref{GaussInt}, the parameter integral over $u$ remains and the final result is expressed in terms of basic elliptic integrals, defined for ${0\leq\ell\leq2k}$,
\begin{align}
I^{\mu_1\dots\mu_{2k}}_{(2k,\ell)}\coloneqq\int_0^\infty\mathrm{ d}u \frac{\hat{g}^{1/2}}{G^{1/2}}u^\ell\tensor*{\left[\operatorname{sym}_{k}\left(G^{-1}\right)\right]}{^{\mu_1\dots\mu_{2k}}}\,.\label{Integrals}
\end{align}
Integrals of the form \eqref{Integrals} occur unavoidably in the multimetric case and are characteristic to the problem. They constitute irreducible structures and can in general not be trivially integrated as for the case of a single metric $\tilde{g}_{\mu\nu}=\hat{g}_{\mu\nu}$.\footnote{An interesting observation is that even for general $\tilde{g}_{\mu\nu}$--in principle--the integrals \eqref{Integrals} can be evaluated explicitly in $d=4$ dimensions. We comment on this in more detail in Appendix \ref{App:Integrals}.}
The evaluation of the trace $T_4$ proceeds in an analogue way. The individual results for $T_2$ and $T_4$ are provided explicitly in Appendix \ref{App:TraceCalc}.

\subsection{Final result: One-loop divergences for the generalized Proca model}
The final result for the divergences of the generalized Proca model in curved spacetime \eqref{ActGenProca} are obtained by adding the standard minimal second order traces \eqref{GenProcMinContributions} and the results for the divergent part of the traces $T_2$ and $T_4$, which are explicitly provided in \eqref{T2Result} and \eqref{T4Result}. Using the integral identities presented in Appendix \ref{App:Integrals}, the number of different integrals \eqref{Integrals} can be reduced. This allows us to represent the final result in the compact form
\begin{widetext}
\begin{align}
\Gamma_{1}^{\mathrm{div}}={}&\frac{1}{32\pi^2\varepsilon}\int{\mathrm{d}}^4x\hat{g}^{1/2}\left[\frac{\hat{\mathcal{G}}}{15}-\frac{1}{60}\left(\tilde{g}^{-1}\right)^{\mu\rho}\left(\tilde{g}^{-1}\right)^{\nu\sigma}\tilde{R}_{\mu\nu}\tilde{R}_{\rho\sigma}-\frac{1}{120}\tilde{R}^2-\frac{1}{5}\hat{R}_{\mu\nu}\hat{R}^{\mu\nu}+\frac{1}{15}\hat{R}^2\right.\nonumber\\
&+\frac{1}{3}\mu^2\tilde{R}+\frac{1}{6}\mu^2\tr \left(\tilde{g}^{-1}\right)\hat{R}-\frac{1}{6}\mu^2\left(\tilde{g}^{-1}\right)^{\mu\nu}\left(\hat{\nabla}_{\alpha}\delta\tensor{\Gamma}{^{\alpha}_{\mu\nu}}+7\hat{R}_{\mu\nu}\right)-\frac{1}{4}\mu^4\tr\left(\tilde{g}^{-2}\right)\nonumber\\
&+\left.C^{(0,0)}I_{(0,0)}
+C^{(2,1)}_{\mu\nu}I_{(2,1)}^{\mu\nu}
+C^{(4,1)}_{\mu\nu\rho\sigma}I_{(4,1)}^{\mu\nu\rho\sigma}
+C^{(4,2)}_{\mu\nu\rho\sigma}I_{(4,2)}^{\mu\nu\rho\sigma}\vphantom{\frac{\hat{\mathcal{G}}}{15}}\right]\,,\label{FinalResult}
\end{align}

\begin{align}
C^{(0,0)}&=\frac{1}{6}\mu^2\left(\tilde{g}^{-2}\right)^{\mu\nu}\tilde{R}_{\mu\nu}-\frac{1}{12}\mu^2\tr\left(\tilde{g}^{-1}\right)\tilde{R}\,,\\
C^{(2,1)}_{\mu\nu}&=\frac{1}{4}\mu^4\left(\tilde{g}^{-2}\right)_{\mu\nu}-\frac{1}{4}\mu^2\left(\tilde{g}^{-1}\right)_{\mu\nu}\left(\frac{1}{2}\mu^2\tr\left(\tilde{g}^{-1}\right)+\frac{1}{3}\hat{R}\right)\nonumber\\
&+\mu^2\left(\tilde{g}^{-1}\right)^{\alpha\beta}\left(-\frac{1}{2}\hat{R}_{\alpha\mu\beta\nu}-\frac{1}{6}\hat{\nabla}_\alpha\tensor{\delta\Gamma}{_\beta_\mu_\nu}+\frac{1}{3}\hat{\nabla}_\beta\tensor{\delta\Gamma}{_\nu_\alpha_\mu}-\frac{1}{3}\hat{\nabla}_\mu\tensor{\delta\Gamma}{_\nu_\alpha_\beta}+\frac{5}{12}\hat{\nabla}_\mu\tensor{\delta\Gamma}{_\alpha_\beta_\nu}-\frac{1}{4}\tensor{\delta\Gamma}{_\alpha_\lambda_\beta}\tensor{\delta\Gamma}{^\lambda_\mu_\nu}\right.\nonumber\\
&\left.+\frac{1}{4}\delta\Gamma^{\lambda}_{\mu\alpha}\delta\Gamma_{\lambda\nu\beta}-\frac{1}{12}\tensor{\delta\Gamma}{_{\alpha\lambda\mu}}\tensor{\delta\Gamma}{^{\lambda}_{\nu\beta}}+\frac{1}{3}\delta\Gamma_{\mu\lambda\alpha}\tensor{\delta\Gamma}{^{\lambda}_{\nu\beta}}-\frac{1}{2}\delta\Gamma_{\mu\lambda\nu}\tensor{\delta\Gamma}{^{\lambda}_{\alpha\beta}}\right)\nonumber\\
&-\frac{1}{6}\mu^2\left(\tilde{g}^{-1}\right)^{\alpha}_{\mu}\left(2\tilde{R}_{\alpha\nu}-5\hat{R}_{\alpha\nu}-\hat{\nabla}_{\lambda}\tensor{\delta\Gamma}{^{\lambda}_{\alpha\nu}}\right)+\frac{1}{12}\mu^2\tr\left(\tilde{g}^{-1}\right)\left(\tensor{\delta\Gamma}{^{\alpha}_{\mu\beta}}\tensor{\delta\Gamma}{^{\beta}_{\alpha\nu}}\right)\,,\\
C^{(4,1)}_{\mu\nu\rho\sigma}&=-\frac{1}{24}\mu^2\tr\left(\tilde{g}^{-1}\right)\hat{\nabla}_{\mu}\delta\Gamma_{\nu\rho\sigma}\,,\\
C^{(4,2)}_{\mu\nu\rho\sigma}&=-\frac{1}{6}\mu^2\hat{\nabla}_{\mu}\delta\Gamma_{\nu\rho\sigma}+\frac{1}{2}\mu^2\delta\Gamma_{\mu\lambda\nu}\tensor{\delta\Gamma}{^{\lambda}_{\rho\sigma}}-\frac{1}{4}\mu^2\delta\Gamma_{\lambda\mu\nu}\tensor{\delta\Gamma}{^{\lambda}_{\rho\sigma}}\,.
\end{align}
\end{widetext}
This constitutes our main result. For compactness, we present the one-loop divergences in terms of the two metrics $\hat{g}_{\mu\nu}$ and $\tilde{g}_{\mu\nu}$. The result in terms of the original metric $g_{\mu\nu}$ and the background mass tensor $M^{\mu\nu}$ can easily be recovered by making use of \eqref{ConfTrafoMetric} and \eqref{DefTildeg}.

As expected on general grounds, the result \eqref{FinalResult} is a local expression, which contains up to four derivatives. The fact that \eqref{FinalResult} is not a polynomial of the invariants of $M^{\mu\nu}$, is related to the role of $M^{\mu\nu}$ as metric in the scalar St\"uckelberg sector.
We emphasize that this result holds for an arbitrary positive definite symmetric background tensor $M^{\mu\nu}$. The original assumption of a strictly positive $M^{\mu\nu}$ is reflected in the result \eqref{FinalResult}, as there is no smooth limit $M^{\mu\nu}\to0$. Note that the result \eqref{FinalResult} is independent of the auxiliary mass scale $\mu$, which can be seen by the invariance of \eqref{FinalResult} under rescaling of $\mu \to\alpha\mu$, with arbitrary constant $\alpha$.

The result has been derived in curved spacetime without considering graviton loops. Nevertheless, the consistency of the renormalization procedure would require to include the induced kinetic terms for $\hat{g}_{\mu\nu}$ an $\tilde{g}_{\mu\nu}$ (for  $g_{\mu\nu}$ and $M^{\mu\nu}$ respectively) in the bare action.
The result \eqref{FinalResult} shows that the essential complexity is not reduced considerably in the limit of a flat spacetime, as the integrals \eqref{Integrals}, associated with the presence of the second metric $\tilde{g}_{\mu\nu}$ (the background mass tensor  $M^{\mu\nu}$) remain. 
The result \eqref{FinalResult} is considerably more complicated than the one-loop result for the nondegenerate vector field. In particular, the result cannot simply be obtained in the limit $\lambda\to0$ from \eqref{eq:EffactNonDegVF}.

For special choices of $M^{\mu\nu}$, the general result \eqref{FinalResult} simplifies and the integrals \eqref{Integrals} can be evaluated explicitly.

\section{Checks and Applications of the generalized Proca model}\label{Sec:ChecksAndAppl}
In this section we reduce our general result \eqref{FinalResult} to several special cases. This provides important cross checks and interesting applications.

\subsection{Trivial index structure}\label{MuCheck}
The simplest case for the general background mass tensor $\tensor{M}{^{\mu\nu}}$, which goes beyond the constant Proca mass term $\tensor{M}{^{\mu\nu}}=m^2\tensor{g}{^{\mu\nu}}$, is the reduction to a spacetime dependent scalar function $X^2(x)$, such that  ${\tensor{M}{^{\mu\nu}}}$ acquires trivial index structure
\begin{align}
M^{\mu\nu}=X^2g^{\mu\nu}\,.\label{Xtrivial}
\end{align} 	
In particular, this includes the case where $X^2$ is proportional to the curvature scalar $R$, which is relevant in cosmological vector field models \cite{Novello1979,Davies1985,Ford1989, Kanno2008,Golovnev2008,Esposito-Farese2010,Heisenberg2014, Belokogne2016}.
In view of \eqref{Xtrivial}, it is easy to see that 
\begin{align}
\tensor{\hat{g}}{_\mu_\nu}={}&\frac{X^2}{\mu^2}\tensor{g}{_\mu_\nu}\,,\label{ConfTrafTriv}\\
\left(\tilde{g}^{-1}\right)^{\mu\nu}={}&\hat{g}^{\mu\nu}\,,\label{tildehatg}\\
\delta\tensor{\Gamma}{^{\lambda}_{\mu\nu}}={}&0\,,\\
\tensor{\left(G^{-1}\right)}{^\mu^\nu}={}&\left(1+u\right)^{-1}\hat{g}^{\mu\nu},\\
\det(\tensor{G}{_\mu_\nu})={}&\left(1+u\right)^4\det\left(\hat{g}_{\mu\nu}\right)\,.\label{DetGTriv}
\end{align}
In this case, the integrals \eqref{Integrals} are trivially evaluated 
\begin{align}
\left.\tensor*{I}{_{(2k,\ell)}^{\mu_1\cdots\mu_{2k}}}\right|_{\tilde{g}=\hat{g}}={}&\tensor*{\left[\operatorname{sym}_{k}\left(\hat{g}\right)\right]}{^{\mu_1\dots\mu_{2k}}}\int_0^\infty\mathrm{ d} u \frac{u^\ell}{(1+u)^{k+2}}\nonumber\\
={}&\frac{(k-\ell)!\ell!}{(k+1)!}\tensor*{\left[\operatorname{sym}_{k}\left(\hat{g}\right)\right]}{^{\mu_1\dots\mu_{2k}}}\,.\label{RedIntTriv}
\end{align}
Inserting \eqref{ConfTrafTriv}--\eqref{DetGTriv} and the explicit results \eqref{RedIntTriv} for the integrals into the general result \eqref{FinalResult}, we obtain the one-loop result in terms of the original metric $g_{\mu\nu}$,
\begin{align}
\Gamma_1^{\mathrm{div}}={}&\frac{1}{32\pi^2\varepsilon}\int\mathrm{d}^4xg^{1/2}\left[\frac{1}{15}\mathcal{G}-\frac{13}{60}\tensor{R}{_\mu_\nu}\tensor{R}{^\mu^\nu}\right.\nonumber\\
&+\frac{7}{120}R^2-\frac{1}{2}RX^2-\frac{3}{2}X^4\nonumber\\
&\left.-\frac{1}{6}R\frac{\Delta X}{X}-3X\Delta X-\frac{1}{2}\left(\frac{\Delta X}{X}\right)^2\right]\,.\label{1LDivGenProcTriv}
\end{align}
As an additional (trivial) cross check of this result, we set $X=m$ in \eqref{1LDivGenProcTriv} and recover the one-loop divergences for the Proca model \eqref{eq:EffactProca}.

An independent way to derive the result \eqref{1LDivGenProcTriv} is to insert \eqref{Xtrivial} directly into the generalized Proca action \eqref{ActGenProca}. 
This leads to the ordinary Proca action \eqref{ActProc}, but with the constant mass $m$ promoted to a spacetime dependent function $X$,
\begin{align}
S[A]=\int\mathrm{d}^4x\,g^{1/2}\left(\frac{1}{4}\tensor{\mathcal{F}}{_\mu_\nu}\tensor{\mathcal{F}}{^\mu^\nu}+\frac{X^2}{2}\tensor{A}{_\mu}\tensor{A}{^\mu}\right)\,.\label{trivAct}
\end{align}
By performing the Weyl transformation \eqref{ConfTrafTriv}, the reduced action \eqref{trivAct} is identical to the Proca action \eqref{ActGenProcConfMet}, but with $g_{\mu\nu}\to\hat{g}_{\mu\nu}$ and $m\to\mu$,
\begin{align}
S[A]=\int\mathrm{d}^4x\,\hat{g}^{1/2}\left(\frac{1}{4}\tensor{\hat{g}}{^\mu^\alpha}\tensor{\hat{g}}{^\nu^\beta}\tensor{\mathcal{F}}{_\mu_\nu}\tensor{\mathcal{F}}{_\alpha_\beta}+\frac{\mu^2}{2}\tensor{\hat{g}}{^\mu^\nu}\tensor{A}{_\mu}\tensor{A}{_\nu}\right)\,.\label{ActTrivIndProc}
\end{align}
The one-loop divergences for \eqref{ActTrivIndProc} are obtained from \eqref{eq:EffactProca}  by performing the inverse Weyl transformation $\hat{g}_{\mu\nu}\to g_{\mu\nu}$ and agree with those obtained from the reduction \eqref{1LDivGenProcTriv} of the general result.

\subsection{Vector field with quartic self interaction}\label{SubSec:SelfIntVecField}
The generalized Proca action for a vector field $\tensor{A}{_\mu}$ with quartic self-interaction is considered in \cite{Barvinsky1989},
\begin{align}
S[A]=\int\mathrm{d}^4x\,g^{1/2}\left[\frac{1}{4}\tensor{\mathcal{F}}{_\mu_\nu}\tensor{\mathcal{F}}{^\mu^\nu}+\frac{\alpha}{4}\left(\tensor{A}{_\mu}\tensor{A}{^\mu}\right)^2\right]\,.\label{ActA4}
\end{align}
The part quadratic in the quantum fluctuation reads
\begin{align}
S_2[\bar{A},\delta A]={}&\int\mathrm{d}^4x\,g^{1/2}\left[\frac{1}{4}\tensor{\mathcal{F}}{_\mu_\nu}(\delta A)\tensor{\mathcal{F}}{^\mu^\nu}(\delta A)\right.\nonumber\\
&\left.+\frac{\alpha}{2}\left(\bar{A}_{\rho}\bar{A}^{\rho}\tensor{g}{^\mu^\nu}+2\tensor{\bar{A}}{^\mu}\tensor{\bar{A}}{^\nu}\right)\tensor{\delta\! A}{_\mu}\tensor{\delta\! A}{_\nu}\right]\,,\label{S2A4}
\end{align}
where the vector field $A_{\mu}$ has been split into background $\bar{A}_{\mu}$ and perturbation $\delta A_{\mu}$,
\begin{align} 
\tensor*{A}{_{\mu}}={}&\tensor*{\bar{A}}{_{\mu}}+\tensor*{\delta A}{_{\mu}},\\
\mathcal{F}(\delta A)={}&\nabla_{\mu}\delta A_{\nu}-\nabla_{\nu}\delta A_{\mu}\,.
\end{align}
In order to establish the connection to the generalized Proca model \eqref{ActGenProca}, we identify the background mass tensor $M^{\mu\nu}$ as
\begin{align}
\tensor{M}{^\mu^\nu}(\bar{A}){}&=\alpha\left(\bar{A}_{\rho}\bar{A}^{\rho}\tensor{g}{^\mu^\nu}+2\tensor{\bar{A}}{^\mu}\tensor{\bar{A}}{^\nu}\right)\,.\label{Mselfint}
\end{align}
According to \eqref{ConfTrafoMetric} and \eqref{DefTildeg}, we have
\begin{align}
\hat{g}_{\mu\nu}={}&\left(g^{\rho\sigma}\xi_{\rho}\xi_{\sigma}\right)g_{\mu\nu}\,,\label{ghatselfint}\\
\tensor{\left(\tilde{g}^{\,-1}\right)}{^\mu^\nu}{}={}&3^{-1/4}\left(\hat{g}^{\mu\nu}+2\xi^{\mu}\xi^{\nu}\right)\,,\label{gtildeselfint}\\
\tensor{\tilde{g}}{_\mu_\nu}={}&3^{1/4}\left(\tensor{\hat{g}}{_\mu_\nu}-\frac{2}{3}\tensor{\xi}{_\mu}\tensor{\xi}{_\nu}\right)\,,\\
\tensor{\delta\Gamma}{^\lambda_\mu_\nu}={}& \frac{2}{3}\left(\tensor{\xi}{_{(\mu}}\hat{\nabla}^\lambda\tensor{\xi}{_{\nu)}}-\tensor{\xi}{_{(\mu}}\tensor{\hat{\nabla}}{_{\nu)}}\tensor{\xi}{^\lambda}-3\tensor{\xi}{^\lambda}\tensor{\hat{\nabla}}{_{(\mu}}\tensor{\xi}{_{\nu)}}\right.\nonumber\\
&\left.\quad+2\tensor{\xi}{^\lambda}\tensor{\xi}{^\alpha}\tensor{\xi}{_{(\mu|}}\tensor{\hat{\nabla}}{_\alpha}\tensor{\xi}{_{|\nu)}}\right)\,.\label{DelGamA4}
\end{align} 
We have defined the normalized vector field
\begin{align}
\xi_{\mu}\coloneqq 3^{1/8}\alpha^{1/2}\frac{\bar{A}_{\mu}}{\mu}\,,\qquad\xi^{\mu}\coloneqq \hat{g}^{\mu\nu}\xi_{\nu}\,,\label{xiselfint}
\end{align}
such that $\xi_{\mu}\xi^{\mu}=1$.
For \eqref{ghatselfint} and \eqref{gtildeselfint}, the integrals \eqref{Integrals} reduce to expressions of the form
\begin{align}
I_{(2k,\ell)}^{\mu_1...\mu_{2k}}=\sum_{n=0}^{k}d_{(2k,\ell)}^{n}\hat{g}^{(\mu_1\mu_2}\cdots\hat{g}^{\mu_{2n-1}\mu_{2n}}\xi^{\mu_{2n+1}}\cdots\xi^{\mu_{2k})}\,.\label{Integralselfint}
\end{align}
We provide a closed form expression for the coefficients $d^{n}_{(2k,\ell)}$ in Appendix \ref{App:Integrals}.
We obtain the divergent part of the one-loop effective action for \eqref{ActA4} from the general result by inserting \eqref{gtildeselfint}--\eqref{DelGamA4} as well as \eqref{Integralselfint} with \eqref{d42} into \eqref{FinalResult},
\begin{widetext}
\begin{align}
\Gamma_1^{\mathrm{div}}={}&\frac{1}{32\pi^2\varepsilon}\int\mathrm{d}^4x\hat{g}^{1/2}\left\{\frac{1}{15}\mathcal{\hat{G}}-\frac{1}{5}\hat{R}_{\mu\nu}\hat{R}^{\mu\nu}+\frac{1}{15}\hat{R}^2-\frac{1}{60}\left(\tilde{g}^{-1}\right)^{\mu\alpha}\left(\tilde{g}^{-1}\right)^{\nu\beta}\tilde{R}_{\mu\nu}\tilde{R}_{\alpha\beta}-\frac{1}{120}\tilde{R}^2-\frac{1}{6}\left(9+2\sqrt{3}\right)\mu^4\right.\nonumber\\
&\left.+\sqrt[4]{3}\mu^2\left[\frac{2}{45}\left(-31+8\sqrt{3}\right)\hat{R}_{\mu\nu}\xi^{\mu}\xi^{\nu}+\frac{1}{18}\left(3-4\sqrt{3}\right)\hat{R}+\frac{1}{45}\left(-31+8\sqrt{3}\right)\left(\hat{\nabla}_{\mu}\xi^{\mu}\right)^2
\right.\right.\nonumber\\
&\left.\left.+\frac{1}{45}\left(-274+136\sqrt{3}\right)\left(\hat{\nabla}_{\mu}\xi_{\rho}\right)\left(\hat{\nabla}_{\nu}\xi^{\rho}\right)\xi^{\mu}\xi^{\nu}+\frac{1}{45}\left(67-36\sqrt{3}\right)\left(\hat{\nabla}_{\mu}\xi_{\nu}\right)\left(\hat{\nabla}^{\mu}\xi^{\nu}\right)\right]\right\}\,.\label{Resultselfint}
\end{align}
\end{widetext}
Instead of $\xi_{\mu}$ and $\tilde{g}_{\mu\nu}$, the authors in \cite{Barvinsky1989} use a different parametrization. The conversion between our result \eqref{Resultselfint} and their result is easily accomplished by \eqref{ghatselfint} and \eqref{xiselfint}. We find that the reduction of our general result to the case of the self-interacting vector field \eqref{Resultselfint} is in agreement with the result obtained in \cite{Barvinsky1989}. This provides a powerful check of our general result \eqref{FinalResult}.

\subsection{Perturbative treatment of the generalized Proca model}
Finally, we test our method by a perturbative calculation, which only relies on the well-established generalized Schwinger-DeWitt technique \cite{Barvinsky1985}.
For this purpose, we assume that the background mass tensor has the form 
\begin{align}
\tensor{M}{^\mu^\nu}=m^2\tensor{g}{^\mu^\nu}+\tensor{Y}{^\mu^\nu},\label{MassPert}
\end{align}
with $Y^{\mu\nu}\ll m^2g^{\mu\nu}$ and perform an expansion in $Y^{\mu\nu}$. 
\subsubsection{Expansion of the general result}
The expansion of the general result \eqref{FinalResult} up to second order reads
\begin{align}
&\Gamma_{1}^{\mathrm{div}}=\Gamma_{1,(0)}^{\mathrm{div}}+\Gamma_{1,(1)}^{\mathrm{div}}+\Gamma_{1,(2)}^{\mathrm{div}}+\mathcal{O}\left(\mathbf{Y}^3\right)\,,\label{GamDivExp}
\end{align}
where $\Gamma_{1,(i)}^{\mathrm{div}}$ is the divergent part of the one-loop effective action, which contains terms of $i$th order in the perturbation $\mathbf{Y}$. The zeroth order $\Gamma_{1,(0)}^{\mathrm{div}}$ is simply given by the one-loop divergences \eqref{eq:EffactProca} for the Proca field.
Before we proceed, let us discuss the structure of the invariants used to represent the result for the higher orders of the perturbative expansion. In $d=4$, the result of any total antisymmetrization among five or more indices is necessarily zero
\begin{align}
\delta_{[\mu}^{\alpha}\delta_{\nu}^{\beta}\delta_{\rho}^{\gamma}\delta_{\sigma}^{\delta}\delta_{\lambda]}^{\omega}\equiv0\,,\label{GDelta}
\end{align}
where the total antisymmetrization is performed with unit weight.
By contracting \eqref{GDelta} with the background tensors $R_{\mu\nu\rho\sigma}$ and $\nabla_{\mu_1}\cdots\nabla_{\mu_{n}}Y^{\rho\sigma}$, we can systematically construct dimensional dependent invariants, which vanish in $d=4$ dimensions. At linear order of the expansion in $\mathbf{Y}$, there is one such invariant
\begin{align}
I_{1}
={}&\delta_{[\mu}^{\alpha}\delta_{\nu}^{\beta}\delta_{\rho}^{\gamma}\delta_{\sigma}^{\delta}\delta_{\lambda]}^{\omega}\tensor{R}{^\mu_\alpha^\nu_\beta}\tensor{R}{^\rho_\gamma^\sigma_\delta}\tensor{Y}{^\lambda_\omega}\nonumber\\
={}&Y^{\mu\nu}\left(\mathcal{G}g_{\mu \nu }-4 R_{\mu }{}^{\gamma \delta \alpha } 	R_{\nu \gamma \delta \alpha }+ 8  R_{\mu \gamma \nu \delta }R^{\gamma \delta }\right.\nonumber\\
&\;\left. + 8 R_{\mu }{}^{\gamma } R_{\nu \gamma } -4 R_{\mu \nu } R\right)\,.\label{DDI}
\end{align}
At quadratic order of the expansion there are two independent dimensional dependent invariants
\begin{align}
I_2={}&\delta_{[\mu}^{\alpha}\delta_{\nu}^{\beta}\delta_{\rho}^{\gamma}\delta_{\sigma}^{\delta}\delta_{\lambda]}^{\omega}\tensor{R}{^\mu_\alpha^\nu_\beta}\nabla_{\gamma}\tensor{Y}{^\rho_\delta}\nabla^{\sigma}\tensor{Y}{^\lambda_\omega}\,,\label{DDI2}\\
I_3={}&\delta_{[\mu}^{\alpha}\delta_{\nu}^{\beta}\delta_{\rho}^{\gamma}\delta_{\sigma}^{\delta}\delta_{\lambda]}^{\omega}\tensor{R}{^\mu_\alpha^\nu_\beta}\tensor{Y}{^\rho_\gamma}\nabla_{\delta}\nabla^{\sigma}\tensor{Y}{^\lambda_\omega}\,.\label{DDI3}
\end{align} 
Since use of \eqref{DDI2} and \eqref{DDI3} does not lead to any simplification of our result, we refrain from presenting the explicit expressions.
For the first order of the expansion $\Gamma_{1,(1)}^{\mathrm{div}}$, linear in $\mathbf{Y}$, we obtain
\begin{align}
\Gamma_{1,(1)}^{\mathrm{div}}={}&\frac{1}{32\pi^2\varepsilon}\int\mathrm{d}^4xg^{1/2}\left\{\frac{1}{12} R Y- \frac{5}{6} R^{\mu \nu } Y_{\mu \nu } - \frac{3}{4} m^2 Y \right.\nonumber\\
 & + \frac{1}{240 m^2}\left[\vphantom{\frac{}{}}- 8R_{\mu \rho \nu \sigma }R^{\mu \nu }  Y^{\rho \sigma } + 2R_{\mu \nu } R^{\mu \nu } Y\right.\nonumber\\
 &\left.\left. - 4R R_{\mu \nu } Y^{\mu \nu }+R^2 Y- 4\left(\Delta R\right)Y \right.\right.\nonumber\\
 &\left.\left.+ 8 \left(\nabla_{\mu }\nabla_{\nu }R\right)Y^{\mu \nu } + 4\left(\Delta R_{\mu \nu }\right)Y^{\mu \nu }\vphantom{\frac{}{}}\right]\vphantom{\frac{3}{4}}\right\}\,,\label{GammaDivLinY}
\end{align}
where we have defined the trace $Y\coloneqq g_{\mu\nu}Y^{\mu\nu}$.
For the second order $\Gamma_{1,(2)}^{\mathrm{div}}$, quadratic in $\mathbf{Y}$, we make use of the tensor algebra bundle \texttt{xAct} \cite{MartinGarcia, Brizuela2009, Nutma2014} and find
\begin{widetext}
\begin{align}
\Gamma_{1,(2)}^{\mathrm{div}}={}&\frac{1}{32\pi^2\varepsilon}\int\mathrm{d}^4xg^{1/2}\left\{
-\frac{1}{8} Y_{\mu \nu } Y^{\mu \nu } - \frac{1}{16} Y^2+\frac{1}{48m^2}\left[\vphantom{\frac{1}{48}}-2 R Y_{\mu \nu } Y^{\mu \nu } + 12 R^{\mu \nu } Y_{\mu }{}^{\rho } Y_{\nu \rho } + R Y^2\right.\right.\nonumber\\
&\left.\left. -4 R^{\mu \nu } Y_{\mu \nu } Y -4 R_{\mu \rho \nu \sigma } Y^{\mu \nu } Y^{\rho \sigma } + 12 Y^{\mu \nu } \nabla_{\nu }\nabla_{\rho }Y_{\mu }{}^{\rho } + 4 Y \nabla_{\rho }\nabla_{\nu }Y^{\nu \rho } +2 Y^{\mu \nu } \Delta Y_{\mu \nu } - Y \Delta Y\vphantom{\frac{1}{48}}\right]\right.\nonumber\\
&\left.+\frac{1}{960m^4}\left[\vphantom{\frac{1}{48}}-2 R_{\mu \nu } R^{\mu \nu } Y^2 - R^2 Y^2 + 8 R_{\mu \nu } R Y Y^{\mu \nu } -2 R^2 Y_{\mu \nu } Y^{\mu \nu } -16 R_{\alpha }{}^{\nu } R_{\nu \mu \rho \sigma } Y^{\alpha \rho } Y^{\mu \sigma } -16 R_{\mu \rho } R^{\mu }{}_{\nu } Y Y^{\nu \rho }\right.\right.\nonumber\\
&\left.\left.+ 8 R_{\mu \nu } R Y^{\mu }{}_{\rho } Y^{\nu \rho } + 32 R^{\mu \nu } R_{\mu \rho \nu \sigma } Y Y^{\rho \sigma } + 24 R_{\mu \rho } R_{\nu \sigma } Y^{\mu \nu } Y^{\rho \sigma } -8 R_{\mu \nu } R_{\rho \sigma } Y^{\mu \nu } Y^{\rho \sigma } -8 R R_{\mu \rho \nu \sigma } Y^{\mu \nu } Y^{\rho \sigma }\right.\right.\nonumber\\
&\left.\left. -16 R_{\mu }{}^{\alpha }{}_{\nu }{}^{\beta } R_{\rho \alpha \sigma \beta } Y^{\mu \nu } Y^{\rho \sigma } -8 R_{\mu \rho } R^{\mu }{}_{\nu } Y^{\nu }{}_{\sigma } Y^{\rho \sigma } -4 R_{\mu \nu } R^{\mu \nu } Y_{\rho \sigma } Y^{\rho \sigma } +16 R_{\mu \rho \nu \sigma } Y^{\mu \nu }\Delta Y^{\rho \sigma }\right.\right.\nonumber\\
&\left.\left. + 32 R_{\mu \sigma \nu \alpha } Y^{\mu \nu } \nabla^{\alpha }\nabla_{\rho }Y^{\rho \sigma } + 32 R_{\nu \rho \sigma \alpha } Y^{\mu \nu } \nabla_{\mu }\nabla^{\alpha }Y^{\rho \sigma } - 8 R_{\nu \rho } Y^{\nu \rho } \Delta Y - 16 R_{\nu \sigma } Y^{\nu \rho }\Delta Y^{\sigma }{}_{\rho }+ 16 R_{\mu \rho \nu \sigma } Y \nabla^{\sigma }\nabla^{\rho }Y^{\mu \nu }\right.\right.\nonumber\\
&\left.\left. + 16 R_{\sigma \nu } Y^{\rho \mu } \nabla_{\mu }\nabla^{\nu }Y^{\sigma }{}_{\rho } -32 R_{\nu \sigma } Y^{\rho \mu } \nabla_{\mu }\nabla_{\rho }Y^{\nu \sigma } + 2 R \left(\nabla_{\mu }Y\right)\nabla^{\mu }Y + 16 R_{\nu \rho } \left(\nabla_{\mu }Y^{\nu \rho }\right) \nabla^{\mu }Y + 24 R \left(\nabla_{\mu }Y\right) \nabla_{\nu }Y^{\mu \nu }\right.\right.\nonumber\\
&\left.\left. -16 R_{\sigma }{}^{\nu } \left(\nabla_{\mu }Y^{\sigma \rho }\right) \nabla_{\nu }Y_{\rho }{}^{\mu } -4 R Y^{\mu \nu } \nabla_{\nu }\nabla_{\mu }Y -12 R Y \nabla_{\nu }\nabla_{\mu }Y^{\mu \nu } -16 R_{\sigma }{}^{\nu } Y^{\sigma \rho } \nabla_{\nu }\nabla_{\mu }Y_{\rho }{}^{\mu } + 16 R_{\mu \nu \rho \sigma } \left(\nabla^{\mu }Y\right) \nabla^{\sigma }Y^{\nu \rho }\right.\right.\nonumber\\
&\left.\left.-8 R_{\mu \rho } Y^{\nu \rho } \nabla_{\nu }\nabla^{\mu }Y + 16 R Y^{\mu \nu } \nabla_{\nu }\nabla_{\rho }Y_{\mu }{}^{\rho } -40 R_{\mu }{}^{\nu } Y \nabla_{\nu }\nabla_{\rho }Y^{\mu \rho } + 32 R_{\mu \rho \nu \sigma } \left(\nabla_{\alpha }Y^{\mu \nu }\right) \nabla^{\sigma }Y^{\alpha \rho }\right.\right.\nonumber\\
&\left.\left. -80 R_{\sigma }{}^{\nu } Y^{\rho \mu } \nabla_{\nu }\nabla^{\sigma }Y_{\rho \mu } + 4 R_{\mu \nu } \left(\nabla^{\mu }Y\right) \nabla^{\nu }Y + 8 R \left(\nabla_{\nu }Y_{\mu }{}^{\rho }\right) \nabla_{\rho }Y^{\mu \nu } + 24 R_{\sigma \mu } \left(\nabla_{\nu }Y^{\sigma \rho }\right) \nabla_{\rho }Y^{\mu \nu } \right.\right.\nonumber\\
&\left.\left. + 40 R_{\nu \sigma } Y^{\nu \rho } \nabla_{\rho }\nabla_{\mu }Y^{\sigma \mu } - 24 R Y^{\mu \nu } \Delta Y_{\mu \nu } - 16 R_{\mu \nu } Y \Delta Y^{\mu \nu } -16 R_{\nu \rho } \left(\nabla_{\mu }Y^{\sigma \mu }\right) \nabla_{\sigma }Y^{\nu \rho }+ 16 R_{\mu \nu } \left(\nabla^{\mu }Y\right) \nabla_{\rho }Y^{\nu \rho } \right.\right.\nonumber\\
&\left.\left. +16 Y \Delta \nabla_{\nu }\nabla_{\mu }Y^{\mu \nu } -4 Y^{\mu \nu } \Delta^2 Y_{\mu \nu } -16 Y^{\mu \nu } \nabla_{\nu }\nabla_{\mu }\nabla_{\sigma }\nabla_{\rho }Y^{\rho \sigma } - 8 Y^{\mu \nu } \nabla_{\nu }\Delta \nabla_{\rho }Y_{\mu }{}^{\rho }-2 Y \Delta^2 Y\vphantom{\frac{1}{48}}\right]
\right\}\,.\label{GammaDivQuadY}
\end{align}	
\end{widetext}	

\subsubsection{Perturbative calculation via the generalized Schwinger-DeWitt technique}

The second order expansion \eqref{GamDivExp} of the general result \eqref{FinalResult} can be checked by a direct perturbative calculation, which only relies on the well-established generalized Schwinger-DeWitt technique, introduced in \cite{Barvinsky1985}. Although conceptually straightforward, the complexity grows rapidly with growing powers of $\mathbf{Y}$ and is already quite involved for the expansion up to second order in $\mathbf{Y}$. Moreover, we are mainly interested in a check of the structures involving derivatives of $\mathbf{Y}$ not tested by the previous checks---apart from the three structures  involving powers of $\Delta X$ that remain in \eqref{1LDivGenProcTriv} after the reduction of $\mathbf{M}$ to the trivial index structure \eqref{Xtrivial}. Therefore, we restrict the direct perturbative calculation to flat spacetime $g_{\mu\nu}=\delta_{\mu\nu}$, $\nabla_{\mu}=\partial_{\mu}$.

Starting point is the action \eqref{ActGenProca} for the generalized Proca field, but with the background mass tensor $\mathbf{M}$ treated perturbatively as in \eqref{MassPert}. The one-loop divergences up to second order in $\mathbf{Y}$ are obtained by expanding \eqref{OpGenProca} around the Proca operator defined in \eqref{ProcOP},
\begin{align}
\mathbf{F}=\left(\mathbf{F}_0+m^2\mathbf{1}\right)+\mathbf{Y}\,.\label{SplitOp}
\end{align}
Using the split \eqref{SplitOp} in the expansion of the logarithm up to second order in $\mathbf{Y}$, we find
\begin{align}
\Tr_1\ln \mathbf{F}={}&\Tr_{1}\ln\left( \mathbf{F}_0+m^2\mathbf{1}+\mathbf{Y}\right)\nonumber\\
={}&\Tr_1\ln \left(\mathbf{F}_0+m^2\mathbf{1}\right)+\Tr_1\left(\mathbf{Y}\frac{\mathbf{1}}{F_{0}+m^2}\right)\nonumber\\
&-\frac{1}{2}\Tr_1 \left(\mathbf{Y}\frac{\mathbf{1}}{F_{0}+m^2}\mathbf{Y}\frac{\mathbf{1}}{F_{0}+m^2}\right)\,.\label{ExpGenProcY2}
\end{align}
Similar to the calculation for the nondegenerate vector field, we use the exact operator identity  \eqref{eq:MassivWard} for the Proca operator ${\mathbf{F}_0+m^2\mathbf{1}}$, which in flat spacetime reads 
\begin{align}
\frac{\delta_{\mu}^{\nu}}{F_0+m^2}=\left(\tensor*{\delta}{_\mu^\nu}-\frac{\partial_{\mu} \partial^{\nu}}{m^2}\right)\frac{1}{-\partial^2+m^2}\,.\label{eq:MassivWardFlat}
\end{align}
Note that the similarity to the expansion \eqref{ExpTRNDVF} for the nondegenerate vector field might be misleading here, as \eqref{ExpGenProcY2} is an expansion in $\mathbf{Y}$, not an expansion in background dimension. This is seen by comparing the counting of background dimension for the two cases. While $\mathbf{Y}=\mathcal{O}\left(\mathfrak{M}^2\right)$, expansion in $\mathbf{Y}$ in \eqref{ExpGenProcY2} is not efficient as  $\mathbf{1}/(F_0+m^2)=\mathcal{O}\left(\mathfrak{M}^{-2}\right)$, such that the combination $\mathbf{Y}\,\mathbf{1}/F_0=\mathcal{O}\left(\mathfrak{M}^0\right)$. Therefore the perturbative series \eqref{ExpGenProcY2} continues up to arbitrary order in $\mathbf{Y}$. 
The counting of background dimension $\mathbf{1}/(F_0+m^2)=\mathcal{O}\left(\mathfrak{M}^{-2}\right)$ can be understood from \eqref{eq:MassivWardFlat}, as
\begin{align}
\frac{1}{-\partial^2+m^2}={}&\mathcal{O}\left(\mathfrak{M}^0\right)\,,\\
 \frac{1}{m^2}\left(-\partial_{\mu}\partial^{\nu}+m^2\delta_{\mu}^{\nu}\right)={}&\mathcal{O}\left(\mathfrak{M}^{-2}\right)\,.
\end{align}
In contrast, in case of the nondegenerate vector field, the analogue expansion \eqref{ExpTRNDVF} in $\mathbf{P}=\mathcal{O}\left(\mathfrak{M}^2\right)$ is efficient in background dimension as $\mathbf{1}/F_\lambda=\mathcal{O}\left(\mathfrak{M}^0\right)$, such that the combination $\mathbf{P}\,\mathbf{1}/F_\lambda=\mathcal{O}\left(\mathfrak{M}^2\right)$ and terms  $\mathcal{O}\left(\mathbf{P}^3\right)$ are already finite.
The difference between the background counting of $\mathbf{1}/\mathbf{F}_{\lambda}$ and $\mathbf{1}/(F_0+m^2)$ can be traced back to the fact that $\Det \mathbf{F}_{\lambda}\neq0$ while $\Det \mathbf{F}_0=0$.

Inserting the identity \eqref{eq:MassivWardFlat} into the expansion \eqref{ExpGenProcY2}, we obtain the following sum of traces up to $\mathcal{O}\left(\mathbf{Y}^2\right)$:
\begin{align}
\Tr_{1}\ln \mathbf{F}={}&\Tr_{1}\ln \left(\mathbf{F}_0+m^2\mathbf{1}\right)\nonumber\\
&+\Tr_{0}\left[B\frac{1}{-\partial^2+m^2}\right]+\Tr_{1}\left[\tensor{Y}{_\mu^\rho}\frac{\delta_\rho^\nu}{-\partial^2+m^2}\right]\nonumber\\
&-\frac{1}{2}\Tr_{1}\left[\tensor{Y}{_\mu^\rho}\frac{\delta_\rho^\sigma}{-\partial^2+m^2}\tensor{Y}{_\sigma^\lambda}\frac{\delta_\lambda^\nu}{-\partial^2+m^2}\right]\nonumber\\
&+\Tr_{1}\left[\tensor{Y}{_\mu^\rho}\frac{\delta_\rho^\sigma}{-\partial^2+m^2}\tensor{Y}{_\sigma^\lambda}\frac{\partial_\lambda\partial^\nu}{m^2}\frac{1 }{-\partial^2+m^2}\right]\nonumber\\
&-\frac{1}{2}\Tr_{0}\left[B\frac{1}{-\partial^2+m^2}B\frac{1}{-\partial^2+m^2}\right].\label{PertGenProcY2Expl}
\end{align}
Here, we have defined the scalar operator 
\begin{align}
B(\partial)\coloneqq\frac{1}{m^2}\partial_{\mu}Y^{\mu\nu}\partial_{\nu}=\mathcal{O}\left(\mathfrak{M}^0\right)\,,
\end{align}
and used the cyclicity of the trace to convert two vector traces in \eqref{PertGenProcY2Expl} into scalar traces. Next we evaluate the divergent contributions of the individual traces in \eqref{PertGenProcY2Expl} separately.
The first trace is just that of the Proca operator \eqref{ProcOP}.
The remaining traces can be further reduced by iterative commutation of all powers of $1/(-\partial^2+m^2)$ to the right, using the identity
\begin{align}
\left[\frac{1}{-\partial^2+m^2},\mathbf{Y}\right]=-\frac{1}{-\partial^2+m^2}\left[-\partial^2,\mathbf{Y}\right]\frac{1}{-\partial^2+m^2}\,.\label{ComInv}
\end{align}
Each iteration of \eqref{ComInv} generates one additional commutator, which increases the number of derivatives acting on the background tensor $\mathbf{Y}$ by at least one,
\begin{align}
\left[-\partial^2,\mathbf{Y}\right]=\left(-\partial^2\mathbf{Y}\right)-2\left(\partial^{\rho}\mathbf{Y}\right)\partial_{\rho}\,.
\end{align}  
In this way the calculation of the divergent part of the trace \eqref{PertGenProcY2Expl} is reduced to the evaluation of a few universal functional traces \cite{Barvinsky1985},
\begin{align}
\mathcal{U}_{\mu_1\dots\mu_p}^{(n,p)}(m^2)=\left.\partial_{\mu_1}\cdots\partial_{\mu_p}\frac{1}{-\partial^2+m^2}\right|_{x^{\prime}=x}^{\mathrm{div}}\,.\label{UFT}
\end{align}
In $d=4$ dimensions, the $\mathcal{U}_{\mu_1\dots\mu_p}^{(n,p)}$'s are divergent for a degree of divergence 
\begin{align}
\chi_{\textrm{div}}=p-2n+4\leq 0.\label{DOD}
\end{align}
Each commutator reduces the number of derivatives $p$ in \eqref{UFT} and therefore decreases the degree of divergence \eqref{DOD} by one. This shows that the iterative procedure \eqref{ComInv} is efficient for the calculation of the divergent contributions.
In flat spacetime, divergences only arise for $\chi_{\mathrm{div}}=k=0,2,4$,
\begin{align}
\mathcal{U}_{\mu_1\dots\mu_{2n+2k}}^{(n,2n+2k)}={}& \frac{(-1)^{n}}{16\pi^2\varepsilon}\frac{2^{2-n-k}m^{2k}}{k!(n-1)!}\tensor*{\left[\operatorname{sym}_{n+k}\left(\delta\right)\right]}{_{\mu_1\dots\mu_{2n+2k}}}.
\end{align}
Following the strategy outlined above, we first calculate the trivial traces which do not involve the evaluation of any commutator:
\begin{align}
\left.\Tr_{1}\ln \left(\mathbf{F}_0+m^2\mathbf{1}\right)\right|^{\mathrm{div}}={}&\frac{1}{16\pi^2\varepsilon}\left(-\frac{3}{2}m^4\right),\label{Tr1Flat}\nonumber\\
\Tr_{0}\left[B\frac{1}{-\partial^2+m^2}\right]^{\mathrm{div}}={}&\frac{1}{16\pi^2\varepsilon}\left(-m^2 Y\right),\nonumber\\
\Tr_{1}\left[\tensor{Y}{_\mu^\rho}\frac{\delta_\rho^\nu}{-\partial^2+m^2}\right]^{\mathrm{div}}={}&\frac{1}{16\pi^2\varepsilon}\left(\frac{1}{4}m^2Y\right)\,,\nonumber\\
\Tr_{1}\left[\tensor{Y}{_\mu^\rho}\frac{\delta_\rho^\sigma}{-\partial^2+m^2}\tensor{Y}{_\sigma^\lambda}\frac{\delta_\lambda^\nu}{-\partial^2+m^2}\right]^{\mathrm{div}}={}&\frac{1}{16\pi^2\varepsilon}Y_{\mu\nu}Y^{\mu\nu}\,.
\end{align}
The remaining two traces require the evaluation of nested commutators. For the first trace we find
\begin{align}
&\Tr_{1}\left[\tensor{Y}{_\mu^\rho}\frac{\delta_\rho^\sigma}{-\partial^2+m^2}\tensor{Y}{_\sigma^\lambda}\frac{\partial_\lambda\partial^\nu }{m^2}\frac{1}{-\partial^2+m^2}\right]^{\mathrm{div}}\nonumber\\
={}&\frac{1}{16\pi^2\varepsilon}\left[\frac{1}{2} Y_{\mu\nu} Y^{\mu\nu} + \frac{1}{3 }Y^{\mu\rho}\frac{ \partial_{\rho}\partial_{\nu}}{m^2}Y_{\mu}{}^{\nu}\right.\nonumber\\
&\left.  + \frac{1}{12 }Y^{\mu\nu} \frac{(-\partial^2)}{m^2}Y_{\mu\nu}\right]\,.
\end{align}
For the second trace we find
\begin{align}
&\Tr_{0}\left[B\frac{1}{-\partial^2+m^2}B\frac{1}{-\partial^2+m^2}\right]^{\mathrm{div}}\nonumber\\
={}&\frac{1}{16\pi^2\varepsilon}\left[\frac{1}{4} Y_{\mu \nu } Y^{\mu \nu } + \frac{1}{8} Y^2  +\frac{1}{12}Y^{\mu \nu } \frac{\left(-\partial^2\right)}{ m^2}Y_{\mu \nu }\right.\nonumber\\
 &+ \frac{1}{6}Y^{\mu \nu } \frac{\partial_{\nu }\partial_{\rho }}{m^2}Y_{\mu }{}^{\rho }+ \frac{1}{24 }Y \frac{\left(-\partial^2\right)}{m^2}Y  - \frac{1}{6 }Y\frac{ \partial_{\mu }\partial_{\nu }}{m^2}Y^{\mu \nu }\nonumber\\
 &+ \frac{1}{30}Y^{\mu \nu }\frac{\partial_{\mu }\partial_{\nu }\partial_{\rho }\partial_{\sigma }}{m^4}Y^{\rho \sigma } + \frac{1}{60}Y^{\mu \nu }\frac{ \partial_{\nu }\left(-\partial^2\right)\partial_{\rho }}{m^4}Y_{\mu }{}^{\rho } \nonumber\\
 & - \frac{1}{30 }Y\frac{\left(-\partial^2\right)\partial_{\mu }\partial_{\nu }}{m^4}Y^{\mu\nu } + \frac{1}{120 }Y^{\mu \nu } \frac{\left(-\partial^2\right)^2}{m^4}Y_{\mu \nu }\nonumber\\
 &\left. + \frac{1}{240 }Y \frac{\left(-\partial^2\right)^2}{m^4}Y\right]\,.\label{Tr6Flat}
\end{align}
Adding the contributions \eqref{Tr1Flat}--\eqref{Tr6Flat} according to \eqref{PertGenProcY2Expl}, we obtain the final result for the one-loop divergences on a flat background up to second order in $Y^{\mu\nu}$,
\begin{widetext}
\begin{align}
\Gamma^{\mathrm{div}}_{1}={}&\frac{1}{32\pi^2\varepsilon}\int\mathrm{d}^4x\left[-\frac{3}{2}m^4-\frac{3}{4}m^2Y-\frac{1}{8} Y_{\mu \nu } Y^{\mu \nu } - \frac{1}{16} Y^2  +\frac{1}{24}Y^{\mu \nu } \frac{\left(-\partial^2\right)}{ m^2}Y_{\mu \nu }+ \frac{1}{4}Y^{\mu \nu } \frac{\partial_{\nu }\partial_{\rho }}{m^2}Y_{\mu }{}^{\rho }-\frac{1}{48 }Y \frac{\left(-\partial^2\right)}{m^2}Y\right.\nonumber\\
& + \frac{1}{12 }Y\frac{ \partial_{\mu}\partial_{\nu }}{m^2}Y^{\mu \nu }
- \frac{1}{60}Y^{\mu \nu }\frac{\partial_{\mu }\partial_{\nu }\partial_{\rho }\partial_{\sigma }}{m^4}Y^{\rho \sigma } - \frac{1}{120}Y^{\mu \nu }\frac{ \partial_{\nu }\left(-\partial^2\right)\partial_{\rho }}{m^4}Y_{\mu }{}^{\rho }  + \frac{1}{60 }Y\frac{\left(-\partial^2\right)\partial_{\mu }\partial_{\nu }}{m^4}Y^{\mu \nu }\nonumber\\
&\left. - \frac{1}{240 }Y^{\mu \nu } \frac{\left(-\partial^2\right)^2}{m^4}Y_{\mu \nu } - \frac{1}{480 }Y \frac{\left(-\partial^2\right)^2}{m^4}Y\right]\,.\label{GamDivSecPert}
\end{align} 
\end{widetext}
The result \eqref{GamDivSecPert} is in perfect agreement with the result for second order expansion \eqref{GammaDivQuadY} of the general result \eqref{FinalResult} reduced to a flat background $g_{\mu\nu}=\delta_{\mu\nu}$.
This provides a powerful check of our general result. In particular, it probes tensorial derivative structures, not captured by the check for the trivial index structure, discussed in Sec. \ref{MuCheck}. It also provides an important independent check of our method, as it has been obtained in a complementary way by a direct application of the generalized Schwinger-DeWitt method. In particular, it neither relies on the St\"uckelberg formalism nor on the bimetric formulation.

\section{Comparison with results in the literature}\label{Sec:ComparisonWithLit}

In this section,  we compare our results with the one-loop divergences of the generalized Proca model \eqref{ActGenProca} obtained previously by different methods and techniques \cite{Toms2015, Buchbinder2017}.

\subsection{Comparison: Local momentum space method}
In \cite{Toms2015}, based on the local momentum space method \cite{Toms2014}, an expression for the one-loop divergences of the generalized Proca model \eqref{ActGenProca} has been obtained. The result [Eq.~(2.24)] in \cite{Toms2015}, includes terms of order $\mathcal{O}\left(R^2,\,RY,\,Y^2\right)$ and, in our conventions, reads
\begin{align}
\Gamma_{1}^{\mathrm{div}}={}&\frac{1}{32\pi^2\varepsilon}\int\mathrm{d}^4x\,g^{1/2}\left(\frac{1}{15}\mathcal{G}-\frac{13}{60}R_{\mu\nu}R^{\mu\nu}+\frac{7}{120}R^2\right.\nonumber\\
&\left.-\frac{1}{2}m^2R-\frac{3}{2}m^4-\frac{5}{6}R_{\mu\nu}Y^{\mu\nu}+\frac{1}{12}RY-\frac{3}{4}m^2Y\right.\nonumber\\
&\left.-\frac{1}{8}Y_{\mu\nu}Y^{\mu\nu}-\frac{1}{16}Y^2\right)\,. \label{GamDivToms}
\end{align}  
The result \eqref{GamDivToms} agrees with the second order expansion \eqref{GamDivExp} of our general result \eqref{FinalResult} under the assumption 
\begin{align}
Y\ll m^2,\qquad R\ll m^2,\qquad \nabla\ll m.\label{PertAss}
\end{align}
In particular, this means that no terms proportional to inverse powers of $m$ appear in the result of \cite{Toms2015} (and therefore, apart from total derivatives, no structures involving derivatives of $Y$). In contrast, our second order expansion \eqref{GamDivExp} was derived only under the assumption $Y\ll m^2$. The coincidence with the terms in \eqref{GamDivToms} therefore provides an independent check of several structures linear and quadratic in $Y$.

\subsection{Comparison: Method of nonlocal field redefinition}
The authors of \cite{Buchbinder2017} have obtained a result for the one-loop divergences of the generalized Proca model \eqref{ActGenProca}, without relying on any perturbative expansion in $Y$ (denoted $X$ in \cite{Buchbinder2017}).
They use the St\"uckelberg formulation to rewrite the generalized Proca theory as a gauge theory for the original Proca field and the St\"uckelberg scalar field. They derive the corresponding block matrix fluctuation operator similar to \eqref{Stueckelberg}. As the authors discuss, with respect to the metric $g_{\mu\nu}$, this operator is both, nonminimal as well as not block-diagonal. The authors perform a ``shiftlike'' transformation of the quantum vector field $A^{\mu}$ [Eq. (22)],
\begin{align}
A^{\mu}=B^{\mu}+\alpha^{\mu\rho}\partial_{\rho}\varphi\,.
\end{align}
Here, $\alpha^{\mu\nu}$ is an a priori undetermined  background tensor.
Next, they derive a condition for which the fluctuation operator is diagonalized [Eq. (25)].
However, in contrast to the statement of the authors, we believe $\alpha^{\mu\nu}$ has to be an operator instead of a background tensor in order to diagonalize the fluctuation operator.
This is critical for the algorithm used in \cite{Buchbinder2017}.
Consequently, our general result \eqref{FinalResult} does not coincide with the one given in {[Eq. (45)]} in \cite{Buchbinder2017}. In  particular, the result of \cite{Buchbinder2017} contains nonlocal structures.

Nevertheless, whether $\alpha^{\mu\nu}$ is an operator or a background tensor is not relevant for the contribution to the one-loop divergences at linear order in $\mathbf{Y}$, as according to [Eq. (25)], $\alpha^{\mu\nu}$ is first order in $\mathbf{Y}$ and therefore the nontrivial contribution in [Eq. (26)] only affects higher orders in $\mathbf{Y}$, starting at $\mathcal{O}\left(\mathbf{Y}^2\right)$.
This explains why the authors reproduce their result with the generalized Schwinger-DeWitt technique \cite{Barvinsky1985} at linear order in $\mathbf{Y}$. The result at linear order in [Eq. (53)] of \cite{Buchbinder2017} is in agreement with our result \eqref{GammaDivLinY} for the approximation linear in $\mathbf{Y}$. Note that this comparison is nontrivial in the sense that their result involves additional terms proportional to the invariant \eqref{DDI}, which vanishes in $d=4$ dimensions.

\section{Conclusions}\label{Sec:Conclusion}
We have investigated the renormalization of generalized vector field models in curved spacetime. We have introduced a classification scheme for different vector field models based on the degeneracy structure of their associated fluctuation operator.
The distinction between different degeneracy classes is partially connected to the nonminimal structures present in the principal symbol of the fluctuation operator.
The simplest theories, where such nonminimal structures can appear, are vector field theories, but these terms are also important for rank two tensor fields, as has been recently discussed for the degenerate fluctuation operator in the context of $f(R)$ gravity \cite{Ruf2018}.

The nondegenerate vector field and the Abelian gauge field are both representatives of two different degeneracy classes, for which the calculation of the one-loop divergences can be performed with standard methods, based on the generalized Schwinger-DeWitt technique \cite{Barvinsky1985}.
We have briefly reviewed these cases and have discussed the technical details associated with the underlying algorithm for the one-loop calculation.
The Proca theory of the massive vector field is the simplest representative in the class of nondegenerate fluctuation operators with a degenerate principal symbol. Only for this special case, standard methods are directly applicable.
In particular, none of these models in the different degeneracy classes can be obtained from one another in a smooth limit---a fact which is related to the discontinuity in the number of propagating degrees of freedom.
Therefore, the different classes have to be studied separately.

The generalized Proca model, which results from the Proca model by generalizing the constant mass term $m^2g^{\mu\nu}$ to a local background mass tensor $M^{\mu\nu}$, is considerably more complicated and can no longer be treated directly by  standard methods. 
Therefore, we have applied the St\"uckelberg formalism in order to reformulate the generalized Proca model as a gauge theory, where the background mass tensor plays a double role as potential in the vector sector and additional metric in the scalar sector of the St\"uckelberg field.
At the price of dealing simultaneously with two metrics, the standard methods are applicable in this case.

Our main result is the derivation of the one-loop divergences for the generalized Proca model \eqref{FinalResult}.
A characteristic feature of this new result is the appearance of the tensorial parameter integrals \eqref{Integrals}. 
The vector field loops induce curvature and $M$-dependent structures. Unless these structures are present in the original action, the generalized Proca model is not perturbatively renormalizable---not even in flat space.
It is interesting that the main complication of the generalized Proca model is not connected to the curved background but originates from the presence of the second metric structure.

We have checked our general result \eqref{FinalResult} by reducing it to simpler models. The one-loop divergences for the trivial index case $M^{\mu\nu}=X^2g^{\mu\nu}$ can be obtained in two ways: by the reduction of the general result \eqref{FinalResult} and independently from the Proca model by a Weyl transformation. We find perfect agreement.
Moreover, to the best of our knowledge, the trivial index case is by itself a genuinely new result.
In addition, we have performed the reduction of our result \eqref{FinalResult} to the case of a vector field with a $(A_{\mu}A^{\mu})^2$ self-interaction. This model has been studied earlier in \cite{Barvinsky1989}. We find perfect agreement.    
Furthermore, we have expanded our general result \eqref{FinalResult} up to second order in the deviation from the Proca model and compared it to a direct perturbative calculation.
We find perfect agreement. As the direct calculation only relies on standard techniques, the agreement does not only provide a powerful check of our general result \eqref{FinalResult}, but also of the method we used to derive it.
Finally, we have compared our results as well as our approach to previous work on the generalized Proca model.
In \cite{Toms2015}, part of our full result for the one-loop divergences \eqref{FinalResult} has been obtained by a different method.
In the corresponding limit, we find that our result reduces to the one derived in \cite{Toms2015}.
In \cite{Buchbinder2017}, yet another method, based on a combination of the St\"uckelberg formalism and a nonlocal field redefinition, has been proposed. However, the result for the one-loop divergences, obtained in \cite{Buchbinder2017}, does not agree with our result \eqref{FinalResult}.
In general, it is quite remarkably that the simple extension from the Proca model to the generalized Proca model leads to such a drastic increase of complexity---already in flat space.

Our result for the one-loop divergences of the generalized Proca model \eqref{FinalResult} with a background mass tensors of the form $M^{\mu\nu}=\zeta_1 R^{\mu\nu}+\zeta_2 R g^{\mu\nu}$ is important for cosmological models, which, at the classical level, have been studied extensively  \cite{Novello1979,Davies1985,Ford1989,Kanno2008,Golovnev2008,Esposito-Farese2010,Heisenberg2014,Belokogne2016}. It would also be interesting to apply the method presented in this paper in the context of massive gravity \cite{Hinterbichler2012, Rham2014}, more general vector field models \cite{Tasinato2014, DeRham2014, Allys2016} and scalar-vector-tensor models \cite{Heisenberg2018, Kase2018}.
\section*{Acknowledgements}
The authors thank I.~L.~Shapiro for correspondence. 
The work of M.~S.~R. is supported by the Alexander von Humboldt Foundation, in the framework of the Sofja Kovalevskaja Award 2014, endowed by the German Federal Ministry of Education and Research.

\appendix
\allowdisplaybreaks[1]
\section{GENERAL FORMALISM}\label{App:GenForm}
\subsection{Heat kernel and one-loop divergences}
For an action functional $S[\bm{\phi}]$ of a general field $\bm{\phi}$ with components $\phi^{A}$, the fluctuation operator $\mathbf{F}(\nabla^x)$, obtained from the second functional derivative has components $\tensor{F}{^A_B}(\nabla^x)=\tensor{\gamma}{^A^C}\tensor{F}{_C_B}(\nabla^x)$, where $\gamma_{AB}$ is a symmetric, nondegenerate and ultralocal bilinear form. 
The Schwinger integral representation of $\mathbf{1}/F$ reads,
\begin{align}
	\frac{\mathbf{1}}{F}=\int_{0}^{\infty}\mathop{}\!\mathrm{d} s\,e^{-s\mathbf{F}}\,,\label{SchwingerGreen}
\end{align}
where $s$ is the ``proper time'' parameter and where we have indicated the bundle structure of inverse operators by the identity matrix $\mathbf{1}$, which has components $\delta_{B}^{A}$.
The Schwinger representation of higher inverse powers $\mathbf{1}/F^n$ with $n\in \mathbb{N}$ and the logarithm of $\mathbf{F}$ are found to be 
\begin{align}
\frac{\mathbf{1}}{F^n}={}&\int_{0}^{\infty}\frac{\mathop{}\!\mathrm{d} s}{(n-1)!}\,s^{n-1}e^{-s\mathbf{F}}\,,
\label{HighPowInvOp}\\
\ln\mathbf{F}={}&-\int_{0}^{\infty}\frac{\mathop{}\!\mathrm{d} s}{s}\,e^{-s\mathbf{F}}\,.
\label{SLog}
\end{align}
The integrand of the proper time integral \eqref{SchwingerGreen} defines the heat kernel 
\begin{align}
	\mathbf{K}(s| x,x^\prime )\coloneqq e^{-s\mathbf{F}}\delta(x,x^\prime )\,.\label{HKF}
\end{align}
With the boundary condition $\mathbf{K}(0|x,x^\prime )=\delta(x,x^\prime )$, it formally satisfies the heat equation
\begin{align}
	\left(\partial_{s}+\mathbf{F}\right)\,\mathbf{K}(s| x,x^\prime )=0\,.
	\label{HeatEq}
\end{align}
For a minimal second order operator
\begin{align}
\mathbf{F}=\mathbf{\Delta}+\mathbf{P}\,,\label{MinSecOp}
\end{align}
a Schwinger-DeWitt representation for the corresponding kernel exists
\begin{align}
	\mathbf{K}(s| x,x^\prime )
	=\frac{g^{1/2}(x^{\prime})}{(4\pi\,s)^{\omega}}\mathfrak{D}^{1/2}(x,x^\prime )\,e^{-\frac{\sigma(x,x^\prime )}{2\,s}}
	\mathbf{\Omega}(s| x,x^\prime )\,,
	\label{DeWittAnK}
\end{align}
with $\omega=d/2$.
The biscalar $\sigma(x,x^\prime )$ is Synge's world function \cite{Synge1960,Poisson2011}, which is defined by 
\begin{align}
	\tensor{\sigma}{^\mu}\tensor{\sigma}{_\mu}=2\sigma\,,\quad\tensor{\sigma}{_\mu}\coloneqq\nabla_{\mu}\sigma,\quad\tensor{\sigma}{^\mu}=\tensor{g}{^\mu^\nu}\nabla_{\nu}\sigma\,.
	\label{defeqsig}
\end{align}
The biscalar ${\mathfrak{D}}(x,x^\prime )$ is the dedensitized Van-Vleck determinant
\begin{align}
	{\mathfrak{D}}(x,x^\prime )=g^{-1/2}(x)\,g^{-1/2}(x^\prime )\,\det\left(\frac{\partial^2\sigma(x,x^\prime )}{\partial \tensor{x}{^\mu}\,\partial \tensor*{x}{^\prime ^{\nu}}}\right)\,,
	\label{VanVleck}
\end{align}
which is defined by the equation
\begin{align}
	{\mathfrak{D}}^{-1}\nabla_{\mu}({\mathfrak{D}}\sigma^{\mu})=2\omega\,.\label{defeqvv}	
\end{align}
All nontrivial physical information is encoded in the matrix-valued bitensor $\mathbf{\Omega}(s |x,x^\prime )$, 
\begin{align}
	\mathbf{\Omega}(s | x,x^\prime )={}&\sum_{n=0}^{\infty}\,s^n\,\mathbf{a}_{n}(x,x^\prime )\,,
	\label{PotAnK}
\end{align}
where the dependence on the proper time parameter $s$ has been explicitly separated by making a power series ansatz with the matrix-valued Schwinger-DeWitt coefficients $\mathbf{a}_{n}(x,x^\prime )$.
Inserting the ansatz (\ref{DeWittAnK}) together with (\ref{PotAnK}) and the minimal second order operator \eqref{MinSecOp} into the heat equation (\ref{HeatEq}), gives a recurrence relation for the Schwinger-DeWitt coefficients 
\begin{align}
	\left[(n+1)+\sigma^{\mu}\nabla_{\mu}\right]\mathbf{a}_{n+1}+\mathfrak{D}^{-1/2}\mathbf{F}\left(\mathfrak{D}^{1/2}\mathbf{a}_{n}\right)=0\,,
	\label{RecSDW}
\end{align}
where $\mathbf{a}_{n}\equiv0$ for $n<0$ implies that $\mathbf{a}_{0}(x,x^\prime )$ satisfies the parallel propagator equation
\begin{align}
	\sigma^{\mu}\nabla_{\mu}\mathbf{a}_{0}(x,x^\prime )=0, \quad \mathbf{a}_{0}(x,x)=\mathbf{1}\,.
	\label{defeqpp}
\end{align}
Therefore, the parallel propagator matrix
\begin{align}
\tensor{\mathcal{P}}{^{A}_{B^{\prime}}}(x,x^{\prime})\coloneqq \tensor{\left[a_0\right]}{^{A}_{B^{\prime}}}(x,x^{\prime})
\end{align}
parallel transports a field $\tensor{\phi}{^{A}}(x)$ at $x$ to a field $\tensor{\left[\mathcal{P}\phi\right]}{^{A'}}(x')$ at $x'$ along the unique geodesic connecting $x$ with $x'$. It only agrees with $\tensor{\phi}{^{A'}}(x')$ if $\sigma^{\mu}\nabla_{\mu}\tensor{\phi}{^{A}}(x)=0$. 
It satisfies
\begin{align}
\tensor{\mathcal{P}}{^{A}_{B^{\prime}}}\tensor{\mathcal{P}}{^{B^{\prime}}_{C}}=\tensor{\delta}{^{A}_{C}}\,.\label{InvParaProp}
\end{align}
Since for a general bitensor, the primed and unprimed indices indicate the corresponding tensorial structure at a given point, the arguments are omitted whenever there is no possibility for confusion.  
The coincidence limits $x^{\prime}\to x$ of the Schwinger-DeWitt coefficients $\mathbf{a}_{n}$ and their derivatives can be obtained recursively. Using dimensional regularization, the Schwinger-DeWitt algorithm gives a closed result for the divergent part of the one-loop effective action of a minimal second order operator \eqref{MinSecOp} for a generic field $\bm{\phi}$. In $d=4$ dimensions, the result is given in terms of the coincidence limit of the second Schwinger-DeWitt coefficient
\begin{align}
	\Gamma_{1}^{\mathrm{div}}={}&\frac{1}{2}\Tr\ln \left(\mathbf{\Delta}+\mathbf{P}\right)\Big|^{\mathrm{div}}\nonumber\\
	={}&-\frac{1}{32\pi^2\varepsilon}\int\mathop{}\!\mathrm{d}^4x\, g^{1/2}\,\tr\,\mathbf{a}_{2}(x,x)\,,\label{1LActionMinimal}\\
		\mathbf{a}_{2}(x,x)={}&\frac{1}{180}\left(\tensor{R}{_{\alpha\beta\gamma\delta}}\tensor{R}{^{\alpha\beta\gamma\delta}}-\tensor{R}{_\alpha_\beta}\tensor{R}{^\alpha^\beta}-6\,\Delta R\right)\mathbf{1}\nonumber\\
	&+\frac{1}{2}\left(\mathbf{P}-\frac{1}{6}R\mathbf{1}\right)^2+\frac{1}{12}\mathbf{R}_{\alpha\beta}\mathbf{R}^{\alpha\beta}+\frac{1}{6}\,\Delta\mathbf{P}\,.\label{SDWa2Coeff}
\end{align}
Here, $1/\varepsilon$ is a pole in dimension $\varepsilon=d/2-2$ and the bundle curvature $\mathbf{R}_{\alpha\beta}$ with components $\tensor{\left[R_{\alpha\beta}\right]}{^{A}_{B}}$ is defined by the commutator
\begin{align}
\left[\nabla_{\mu},\nabla_{\nu}\right]\phi^{A}=\tensor{\left[R_{\mu\nu}\right]}{^A_B}\phi^{B}\,.
\end{align}
In particular, for a scalar and vector field, we have
\begin{align}
\left[\nabla_{\mu},\nabla_{\nu}\right]\varphi=0,\quad\left[\nabla_{\mu},\nabla_{\nu}\right]A_{\rho}=\tensor{R}{_{\mu\nu\rho}^{\sigma}}A_{\sigma}\,.
\end{align}

\subsection{Covariant Taylor expansion and Synge's rule}
The covariant Taylor expansion of a scalar function $f(x)$ around $x^{\prime}=x$ is given by \cite{Barvinsky1985},
\begin{align}
f(x^{\prime})={}&\sum_{k=0}^{\infty}\frac{(-1)^k}{k!}\left[\nabla_{\mu_1^{\prime}}\cdots\nabla_{\mu_{k}^{\prime}}f(x^{\prime})\right]_{x^{\prime}=x}\sigma^{\mu_1}\cdots\sigma^{\mu_k}\,.\label{CovTaylorScal}
\end{align}
The generalization of this expansion for fields $\phi^{A}(x)$ requires use of the parallel propagator $\tensor{\mathcal{P}}{^{A}_{B^{\prime}}}$,
\begin{align}
\tensor{\left[\mathcal{P}\phi\right]}{^{A^{\prime\prime}}}=\tensor{\mathcal{P}}{^{A^{\prime\prime}}_{B^{\prime}}}\tensor{\phi}{^{B^{\prime}}}\,.\label{BiTParallel}
\end{align}
The right-hand side of \eqref{BiTParallel} transforms as scalar at $x^{\prime}$. Therefore, we can consider the right-hand side as a scalar function of $x^{\prime}$ and apply the covariant Taylor expansion \eqref{CovTaylorScal} around $x^{\prime}=x$,
\begin{align}
	\tensor{\mathcal{P}}{^{A^{\prime\prime}}_{B^{\prime}}}\tensor{\phi}{^{B^{\prime}}}
	={}&\sum_{k=0}^{\infty}\frac{(-1)^k}{k!}\left[\nabla_{\mu_1^{\prime}}\cdots\nabla_{\mu_{k}^{\prime}}\tensor{\mathcal{P}}{^{A^{\prime\prime}}_{B^{\prime}}}\tensor{\phi}{^{B^{\prime}}}\right]_{x^{\prime}=x}\nonumber\\
	&\times \sigma^{\mu_1}\cdots\sigma^{\mu_k}\,.\label{CovTaylorPP}
\end{align}
Using that the coincide limits of the totally symmetrized covariant derivatives acting on the parallel propagator are zero, 
\begin{align}
\left[\nabla_{(\mu^{\prime}_1}...\nabla_{\mu^{\prime}_{k})}\tensor{\mathcal{P}}{^{A}_{B^{\prime}}}\right]_{x^{\prime}=x}=0\,,\quad k>0\,,
\end{align} 
the parallel propagator can be freely commuted though the derivatives in \eqref{CovTaylorPP}.
Further setting $x^{\prime\prime}$ to $x^{\prime}$ and making use of \eqref{defeqpp}, we find 
\begin{align}
\phi^{A^{\prime}}={}&\tensor{\mathcal{P}}{^{A^{\prime}}_{B}}\sum_{k=0}^{\infty}\frac{(-1)^k}{k!}\left[\nabla_{\mu_1^{\prime}}\cdots\nabla_{\mu_{k}^{\prime}}\tensor{\phi}{^{B^{\prime}}}\right]_{x^{\prime}=x}\nonumber\\
&\times \sigma^{\mu_1}\cdots\sigma^{\mu_k}\,.
\end{align}
Applying this expansion to a general bitensor $\tensor{T}{^{A}_{B^{\prime}}}$ around coincidence $x=x^{\prime}$, shows that all the information of  $\tensor{T}{^{A^{\prime}}_{B}}$ is contained in the coincidence limits of its derivatives
\begin{align}
\tensor{T}{^{A^{\prime}}_{B}}={}&\tensor{\mathcal{P}}{^{A^{\prime}}_{C}}\sum_{k=0}^{\infty}\frac{(-1)^k}{k!}\left[\nabla_{\mu_1^{\prime}}\cdots\nabla_{\mu_{k}^{\prime}}\tensor{T}{^{C^{\prime}}_{B}}\right]_{x^{\prime}=x}\nonumber\\
&\times \sigma^{\mu_1}\cdots\sigma^{\mu_k}\,.\label{CovTaylorBitensor}
\end{align}

\subsection{Coincidence limits}
The coincidence limits for $\sigma$, $\mathfrak{D}^{1/2}$, $\mathbf{a}_{n}$ as well as derivatives thereof can be obtained recursively by repeatedly taking derivatives of the defining equations \eqref{defeqsig}, \eqref{defeqvv}, \eqref{RecSDW} and \eqref{defeqpp}. Commutation of covariant derivatives to a canonical order induces curvature terms. Inserting the coincidence limits from lower orders of the recursion, higher coincidence limits are obtained systematically \cite{DeWitt1965}.
In case a bitensor involves derivatives at different points, we can recursively reduce the coincidence limits of primed derivatives to coincidence limits involving only unprimed derivatives by Synge's rule \cite{Synge1960, Poisson2011},
\begin{align}
[\nabla_{\mu^{\prime}}\tensor{T}{}]_{x^{\prime}=x}=\nabla_{\mu}[\tensor{T}{}]_{x^{\prime}=x}-[\nabla_{\mu}\tensor{T}{}]_{x^{\prime}=x}\,.
\end{align}
Here, $T$ represents an arbitrary bitensor. The first few coincidence limits of $\sigma$, $\mathfrak{D}^{1/2}$ and $\mathbf{a}_n$ are easily obtained. In this article, apart from the coincidence limit for $\mathbf{a}_2=\mathcal{O}\left(\mathfrak{M}^{4}\right)$, provided already in \eqref{SDWa2Coeff}, we only need coincidence limits up to ${\cal O}\left(\mathfrak{M}^2\right)$.
Note that, when performing the covariant Taylor expansion \eqref{CovTaylorScal}, we have to specify the metric with respect to which we perform the covariant Taylor expansion, as the metric enters the world function $\sigma$ and the definition of the covariant derivative $\nabla_{\mu}$. 
For a metric $\mathring{g}_{\mu\nu}$, not necessarily compatible with the connection $\nabla_{\mu}$, it is natural to introduce the tensor which measures the difference between the connection $\nabla_{\mu}$ and the Levi-Civita connection associated with $\mathring{g}_{\mu\nu}$,
\begin{align}
 \tensor{\mathring{\delta\Gamma}}{^{\rho}_{\mu\nu}}=\frac{1}{2}\tensor{\left(\mathring{g}^{-1}\right)}{^{\rho\alpha}}\left(\tensor{\nabla}{_\mu}\tensor{\mathring{g}}{_{\alpha\nu}}+\tensor{\nabla}{_\nu}\tensor{\mathring{g}}{_{\mu\alpha}}-\tensor{\nabla}{_\alpha}\tensor{\mathring{g}}{_{\mu\nu}}\right)\,.
\end{align}
We provide the coincidence limits for the world function $\mathring{\sigma}$ and Van-Vleck biscalar $\mathring{\mathfrak{D}}^{1/2}$ as well as derivatives thereof  up to ${\cal O}\left(\mathfrak{M}^2\right)$,
\begin{align}
\left[\mathring{\sigma}\right]_{x'=x}={}&0\,,\label{ColimSigT1}\\
\left[\nabla_{\mu}\mathring{\sigma}\right]_{x'=x}={}&0\,,\\
\left[\nabla_{\mu}\nabla_{\nu}\mathring{\sigma}\right]_{x'=x}={}&\tensor{\mathring{g}}{_{\mu\nu}}\,,\\
\left[\nabla_{\mu}\nabla_{\nu}\nabla_{\rho}\mathring{\sigma}\right]_{x'=x}={}&\mathring{\delta\Gamma}^{\alpha}_ {\mu\nu}\tensor{\mathring{g}}{_{\alpha\rho}}+\mathring{\delta\Gamma}^{\alpha}_ {\mu\rho}\tensor{\mathring{g}}{_{\nu\alpha}}+\mathring{\delta\Gamma}^{\alpha}_ {\nu\rho}\tensor{\mathring{g}}{_{\mu\alpha}}\nonumber\\
={}&3\nabla_{(\mu}\mathring{g}_{\nu\rho)}\,,\\
\left[\nabla_{\mu}\nabla_{\nu}\nabla_{\rho}\nabla_{\sigma}\mathring{\sigma}\right]_{x'=x}={}&-\frac{2}{3}\tensor{\mathring{R}}{_{\mu(\rho|\nu|\sigma)}}+3\tensor{\mathring{g}}{_{\alpha\beta}}\mathring{\delta\Gamma}^{\alpha}_{\mu(\sigma}\mathring{\delta\Gamma}^{\beta}_{\nu\rho)}\nonumber\\
&+2\mathring{g}_{\alpha(\rho|}\left(\nabla_{\mu}\mathring{\delta\Gamma}^{\alpha}_{\nu|\sigma)}+\mathring{\delta\Gamma}^{\alpha}_{\mu\beta}\mathring{\delta\Gamma}^{\beta}_{\nu|\sigma)}\right)\nonumber\\
&+2\mathring{g}_{\alpha(\mu}\left(\nabla_{\nu)}\mathring{\delta\Gamma}^{\alpha}_{\rho\sigma}+\mathring{\delta\Gamma}^{\alpha}_{\nu)\beta}\mathring{\delta\Gamma}^{\beta}_{\rho\sigma}\right)\,,\\
\left[\nabla_{\mu}\mathring{\mathfrak{D}}^{1/2}\right]_{x'=x}={}&0\,,\\
\left[\nabla_{\mu}\nabla_{\nu}\mathring{\mathfrak{D}}^{1/2}\right]_{x'=x}={}&\frac{1}{6}\tensor{\mathring{R}}{_{\mu\nu}}\label{ColimVVT1}\,.
\end{align}
The corresponding coincidence limits of the Schwinger-DeWitt coefficients read
\begin{align}
\left[\mathring{\mathbf{a}}_0\right]_{x^{\prime}=x}={}&\mathbf{1}\,,\\
\left[\nabla_{\alpha}\mathring{\mathbf{a}}_{0}\right]_{x^{\prime}=x}={}&0\,,\\
\left[\nabla_{\alpha}\nabla_{\beta}\mathring{\mathbf{a}}_{0}\right]_{x^{\prime}=x}={}&\frac{1}{2}\mathbf{R}_{\alpha\beta}\,,\\
\left[\mathring{\mathbf{a}}_{1}\right]_{x^{\prime}=x}={}&\frac{1}{6}\mathring{R}\mathbf{1}-\mathbf{P}\,.\label{Colima1Ring}
\end{align}
In case the metric $\mathring{g}_{\mu\nu}$ is compatible with the connection $\nabla_{\mu}\mathring{g}_{\nu\rho}=0$, the coincidence limits \eqref{ColimSigT1}--\eqref{Colima1Ring} reduce to the well-known results \cite{Barvinsky1985}.
\onecolumngrid

\section{DETAILS OF THE CALCULATION FOR THE NONDEGENERATE VECTOR FIELD}\label{App:NDVT}
The traces in \eqref{ExpTRNDVF} can be systematically reduced to the evaluation of tabulated universal functional traces
\begin{align}
\left.\Tr_{1}\ln\mathbf{F}_\lambda\right|^{\mathrm{div}}
={}&\Tr_{1}\ln\left.\mathbf{\Delta}_{\mathrm{H}}\right|^{\mathrm{div}},\,\label{TracesNDVF}	\\
\left.\Tr_{1}\left(\mathbf{P}\frac{\mathbf{1}}{F_{\lambda}}\right)\right|^{\mathrm{div}}={}&\left.\tensor{P}{_{\mu}^{\nu}}\frac{\delta^{\nu}_{\mu}}{\Delta_{\mathrm{H}}}\right|_{x^{\prime}=x}^{\mathrm{div}}-\gamma \left.\tensor{P}{^{\mu}^{\nu}}\nabla_{\mu}\nabla_{\nu}\frac{1}{\Delta^2}\right|_{x^{\prime}=x}^{\mathrm{div}}\,,\\
\left.\Tr_{1}\left(\mathbf{P}\frac{\mathbf{1}}{F_{\lambda}}\mathbf{P}\frac{\mathbf{1}}{F_{\lambda}}\right)\right|^{\mathrm{div}}={}&\left.\tensor{P}{_{\mu}^{\nu}}\tensor{P}{_{\nu}^{\mu}}\frac{1}{\Delta^2}\right|_{x^{\prime}=x}^{\mathrm{div}}-2\gamma\left.\tensor{P}{^{\nu}^{\rho}}\tensor{P}{_{\rho}^{\mu}}\nabla_{\mu}\nabla_{\nu}\frac{1}{\Delta^3}\right|_{x^{\prime}=x}^{\mathrm{div}}+\gamma^2\left.\tensor{P}{^{\mu}^{\nu}}\tensor{P}{^{\rho}^{\sigma}}\nabla_{\mu}\nabla_{\nu}\nabla_{\rho}\nabla_{\sigma}\frac{1}{\Delta^4}\right|_{x^{\prime}=x}^{\mathrm{div}}\,.\label{Tr3}
\end{align}
In \eqref{TracesNDVF}, we have used the definition of the Hodge operator \eqref{Hodge} along with the identity \eqref{NDVFOpID} and the fact that
\begin{align}
\Tr_{1}\ln\left[1+\lambda\nabla_{\mu}\frac{1}{\Delta}\nabla^{\nu}\right]^{\mathrm{div}}=0,
\end{align}
which is formally seen by expanding the logarithm, making use of the cyclicity of the trace and resumming the terms. The traces in \eqref{Tr3} are already $\mathcal{O}\left(\mathfrak{M}^4\right)$, which allows to freely commute all operators and use
\begin{align}
\frac{\delta^{\mu}_{\nu}}{\Delta_{\mathrm{H}}}=\delta^{\mu}_{\nu}\frac{1}{\Delta}+\mathcal{O}\left(\mathfrak{M}\right).
\end{align}
The logarithmic trace \eqref{TracesNDVF} is evaluated directly with \eqref{1LActionMinimal}, while for the remaining traces we use the following universal functional traces
\begin{align}
\left.\frac{\delta^{\nu}_{\mu}}{\Delta_{\mathrm{H}}}\right|_{x^{\prime}=x}^{\mathrm{div}}={}&\frac{g^{1/2}}{16\pi^2\varepsilon}\left(\frac{1}{6}R\,\delta^{\nu}_{\mu}-\tensor{R}{_{\mu}^{\nu}}\right)\,,\label{UFT1}\\
\left.\nabla_{\mu}\nabla_{\nu}\frac{1}{\Delta^2}\right|_{x^{\prime}=x}^{\mathrm{div}}={}&\frac{g^{1/2}}{16\pi^2\varepsilon}\left(\frac{1}{6}R_{\mu\nu}-\frac{1}{12}Rg_{\mu\nu}\right),\\
\left.\nabla_{\mu_1}\cdots\nabla_{\mu_{2n-4}}\frac{1}{\Delta^{n}}\right|_{x^{\prime}=x}^{\mathrm{div}}={}&\frac{g^{1/2}}{16\pi^2\varepsilon}\frac{(-1)^{n}}{2^{n-2}(n-1)!}\left[\operatorname{sym}_{n-2}\left(g\right)\right]_{\mu_1\dots\mu_{2n-4}}\label{UFT3}\,.
\end{align}
Inserting \eqref{UFT1}--\eqref{UFT3} into \eqref{TracesNDVF}--\eqref{Tr3}, we find
\begin{align}
\left.\Tr_{1}\ln\mathbf{F}_\lambda\right|^{\mathrm{div}}={}&\frac{1}{16\pi^2\varepsilon}\int\mathrm{d}^4xg^{1/2}\left[\frac{11}{180}\mathcal{G}-\frac{7}{30}R_{\mu\nu}R^{\mu\nu}+\frac{1}{20}R^2\right]\,,\\
\left.\Tr_{1}\left(\mathbf{P}\frac{\mathbf{1}}{F_{\lambda}}\right)\right|^{\mathrm{div}}={}&\frac{1}{16\pi^2\varepsilon}\int\mathrm{d}^4xg^{1/2}\left[\left(\frac{1}{6}+\frac{\gamma}{12}\right)RP-\left(1+\frac{\gamma}{6}\right)R_{\mu\nu}P^{\mu\nu}\right]\,,\\
\left.\Tr_{1}\left(\mathbf{P}\frac{\mathbf{1}}{F_{\lambda}}\mathbf{P}\frac{\mathbf{1}}{F_{\lambda}}\right)\right|^{\mathrm{div}}={}&\frac{1}{16\pi^2\varepsilon}\int\mathrm{d}^4xg^{1/2}\left[\left(1+\frac{\gamma}{2}+\frac{\gamma^2}{12}\right)P_{\mu\nu}P^{\mu\nu}+\frac{\gamma^2}{24}P^2\right]\,.\label{TracesNDVFExpl}
\end{align}
Adding all traces according to \eqref{ExpTRNDVF}, we obtain the final result \eqref{eq:EffactNonDegVF}.

\section{MULTIPROPAGATOR BIMETRIC TRACES OF THE GENERALIZED PROCA MODEL}\label{App:TraceCalc}

\subsection{Divergent part of the second order trace}

The evaluation of the second order trace $T_2$ constitutes the most complex part of the calculation.
The functional trace $T_2$ is divergent and needs to be regularized. We use regularization in the dimension $d=2\omega$. Explicitly, the functional trace of the convoluted integral kernels is given by
\begin{align}
T_2={}&\Tr_0 \left(\tensor{\Pi}{^\alpha}\frac{\tensor*{\delta}{_\alpha^\beta}}{D_1}\tensor*{\Pi}{^\dagger_\beta}\frac{1}{D_0}\right)=\int\mathrm{d}^{2\omega}x\,\mathrm{d}^{2\omega}x^{\prime}\left[\Sigma^{\beta^\prime}_1(x,x^\prime)\Sigma_{\beta^\prime}^2(x^\prime,x)\right]\,,\label{T2convolution}
\end{align}
where the kernels $\Sigma^{\beta^\prime}_1(x,x^\prime)$ and $\Sigma_{\beta^\prime}^2(x^\prime,x)$ are defined as
\begin{align}
	\Sigma^{\beta^\prime}_1(x,x^\prime)\coloneqq
	\Pi^{\alpha}\frac{\delta^{\beta^{\prime}}_\alpha}{D_1}\hat{\delta}(x,x^{\prime})\,,\qquad
	\Sigma_{\beta^\prime}^2(x^\prime,x)\coloneqq \Pi^{\dagger}_{\beta'}\frac{1}{D_0}\tilde{\delta}(x^{\prime},x)\label{SigmaKern}\,.
\end{align}
First, we insert the integral representations \eqref{SchwingerGreen}, \eqref{HKF} and \eqref{DeWittAnK} for the kernels of the inverse propagators 
\begin{align}
\frac{\delta^{\mu^{\prime}}_\nu}{D_1}\hat{\delta}(x,x^{\prime})
={}&\int_0^\infty\frac{\mathrm{ d}s}{(4\pi s)^{\omega}} e^{-\frac{\hat{\sigma}(x,x^{\prime})}{2s}}\hat{\mathfrak{D}}^{1/2}(x,x^{\prime})\,\tensor{\hat{\Omega}}{_\nu^{\mu^{\prime}}}(s|x,x^{\prime})\hat{g}^{1/2}(x^{\prime})\,,\label{InvVec}\\
\frac{1}{D_0}\tilde{\delta}(x^{\prime},x)={}&\int_0^\infty\frac{\mathrm{ d}t}{(4\pi t)^{\omega}} e^{-\frac{\tilde{\sigma}(x^{\prime},x)}{2t}}\tilde{\mathfrak{D}}^{1/2}(x^{\prime},x)\,\tilde{\Omega}_0(t|x^{\prime},x)\tilde{g}^{1/2}(x)\label{InvScal}\,,
\end{align}
together with the explicit expressions \eqref{VOpComp} for $\Pi^{\alpha}$ and $\Pi^{\dagger}_{\mu'}$ into \eqref{SigmaKern}. Then, we expand the derivatives in each factor in \eqref{SigmaKern} according to the Leibniz rule and collect terms up to $\mathcal{O}\left(\mathfrak{M}^3\right)$,
\begin{align}
\Sigma^{\beta^\prime}_1(x,x^\prime)={}&\mu\hat{\nabla}_\sigma\tensor{\left(\tilde{g}^{\,-1}\right)}{^\sigma^\alpha}\frac{\delta^{\beta^{\prime}}_\alpha}{D_1}\hat{\delta}(x,x^{\prime})\nonumber\\
={}&\mu\int_0^{\infty}\frac{\mathrm{d}s}{(4\pi s)^\omega}\hat{g}^{1/2} e^{-\frac{\hat{\sigma}}{2s}}\mathfrak{\hat{D}}^{1/2}\left\{\left[
\hat{\nabla}_\sigma\tensor{\left(\tilde{g}^{\,-1}\right)}{^\sigma^\alpha}+
\tensor{\left(\tilde{g}^{\,-1}\right)}{^\sigma^\alpha}\mathfrak{\hat{D}}^{-1/2}\left(\hat{\nabla}_\sigma\mathfrak{\hat{D}}^{1/2}\right)-\frac{1}{2s}\hat{\sigma}_\sigma\tensor{\left(\tilde{g}^{\,-1}\right)}{^\sigma^\alpha}\right]\tensor{\mathcal{P}}{_{\alpha}^{\beta^{\prime}}} \right.\nonumber\\
&\left.
+
\tensor{\left(\tilde{g}^{\,-1}\right)}{^\sigma^\alpha}\left(\hat{\nabla}_\sigma\tensor{\mathcal{P}}{_{\alpha}^{\beta^{\prime}}}\right)-\frac{1}{2}\hat{\sigma}_\sigma\tensor{\left(\tilde{g}^{\,-1}\right)}{^\sigma^\alpha}\tensor{\left[a_1\right]}{_{\alpha}^{\beta^{\prime}}}\right\}+\mathcal{O}\left(\mathfrak{M}^4\right)\,,\label{Factor1}\\
\Sigma_{\beta^\prime}^2(x^\prime,x)={}&-\mu\tensor{\left(\tilde{g}^{-1}\right)}{_{\beta^{\prime}}^{\sigma^{\prime}}}\hat{\nabla}_{\sigma^{\prime}}\frac{1}{D_0}\tilde{\delta}(x^{\prime},x)\nonumber\\
={}&-\mu\int_0^{\infty}\frac{\mathrm{d}t}{(4\pi t)^\omega}\tilde{g}^{1/2}\tilde{\mathfrak{D}}^{1/2}e^{-\frac{\tilde{\sigma}}{2t}}\left\{
\tensor{\left(\tilde{g}^{-1}\right)}{_{\beta^{\prime}}^{\sigma^{\prime}}}\hat{\nabla}_{\sigma^{\prime}}\tilde{a}_0-\frac{1}{2t}
\tensor{\left(\tilde{g}^{-1}\right)}{_{\beta^{\prime}}^{\sigma^{\prime}}}\tilde{\sigma}_{\sigma^{\prime}}\tilde{a}_0+
\tensor{\left(\tilde{g}^{-1}\right)}{_{\beta^{\prime}}^{\sigma^{\prime}}}\tilde{\mathfrak{D}}^{-1/2}\hat{\nabla}_{\sigma^{\prime}}\tilde{\mathfrak{D}}^{1/2}\tilde{a}_0\right.\nonumber\\
&\left.-
\tensor{\left(\tilde{g}^{-1}\right)}{_{\beta^{\prime}}^{\sigma^{\prime}}}\tilde{\sigma}_{\sigma^{\prime}}\tilde{a}_1\right\}+\mathcal{O}\left(\mathfrak{M}^4\right)\,.\label{Factor2}
\end{align}
Next, we apply the covariant Taylor expansion \eqref{CovTaylorBitensor} separately to the terms in the curly brackets in \eqref{Factor1} and \eqref{Factor2} up to terms with background dimension ${\cal O}\left(\mathfrak{M}^2\right)$. This requires knowledge of the covariant Taylor expansion of the basic geometrical bitensors up to $\mathcal{O}\left(\mathfrak{M}^2\right)$,
\begin{align}
\tilde{\sigma}={}&\tensor{\tilde{g}}{_\alpha_\beta}\hat{\sigma}^\alpha\hat{\sigma}^\beta
-\frac{1}{2}\left(\tensor{\tilde{g}}{_\alpha_\lambda}\tensor{\delta\Gamma}{^\lambda_\beta_\gamma}\right)\hat{\sigma}^\alpha\hat{\sigma}^\beta\hat{\sigma}^\gamma\nonumber\\
&+\frac{1}{24}\left(4\tensor{\tilde{g}}{_\alpha_\nu}\tensor{\delta\Gamma}{^\nu_\beta_\lambda}\tensor{\delta\Gamma}{^\lambda_\gamma_\delta}+3\tensor{\tilde{g}}{_\nu_\lambda}\tensor{\delta\Gamma}{^\nu_\alpha_\beta}\tensor{\delta\Gamma}{^\lambda_\gamma_\delta}+4\tensor{\tilde{g}}{_\alpha_\lambda}\hat{\nabla}_\beta\tensor{\delta\Gamma}{^\lambda_\gamma_\delta}\right)\hat{\sigma}^\alpha\hat{\sigma}^\beta\hat{\sigma}^\gamma\hat{\sigma}^\delta\,,\label{CovTaylsig}\\
\tensor{\left(\tilde{g}^{-1}\right)}{_{\beta^\prime}^{\rho^{\prime}}}\tilde{\sigma}_{\rho^\prime}={}&\tensor{\mathcal{P}}{_{\beta^\prime}^{\rho}}{}\left[-\hat{\sigma}_\rho-\frac{1}{2}\tensor{\delta\Gamma}{_\rho_\alpha_\beta}\hat{\sigma}^\alpha\hat{\sigma}^\beta
-\frac{1}{6}\left(\tensor{\delta\Gamma}{_\rho_\alpha_\lambda}\tensor{\delta\Gamma}{^\lambda_\beta_\gamma}-2\hat{\nabla}_\gamma\tensor{\delta\Gamma}{_\rho_\alpha_\beta}\right)\hat{\sigma}^\alpha\hat{\sigma}^\beta\hat{\sigma}^\gamma\right]\,,\\
\hat{\mathfrak{D}}^{1/2}={}&1+\frac{1}{12}\tensor{\hat{R}}{_\alpha_\beta}\hat{\sigma}^\alpha\hat{\sigma}^\beta\,,\\
\tilde{\mathfrak{D}}^{1/2}={}&1+\frac{1}{12}\tensor{\tilde{R}}{_\alpha_\beta}\hat{\sigma}^\alpha\hat{\sigma}^\beta\,,\\
\tensor{\left(\tilde{g}^{-1}\right)}{_{\beta^\prime}^{\sigma^\prime}}\tensor*{\tilde{\nabla}}{_{\rho^\prime}}
\tilde{\mathfrak{D}}^{1/2}={}&-\frac{1}{6}\tensor{\mathcal{P}}{_{\beta^\prime}^\rho}\tensor{\left(\tilde{g}^{-1}\right)}{_{\rho}^{\alpha}}\tilde{R}_{\alpha\beta}\sigma^\beta\,,\\
\hat{\nabla}_\alpha\hat{\mathfrak{D}}^{1/2}={}&\frac{1}{6}\tensor{\hat{R}}{_\alpha_\beta}\hat{\sigma}^\beta\,,\\
\left(\hat{\nabla}_\sigma\tensor{\mathcal{P}}{_{\alpha}^{\beta^\prime}}\right)={}&-\frac{1}{2}\tensor{\mathcal{P}}{_{\lambda}^{\beta^\prime}}\tensor{\hat{R}}{_\sigma_\gamma_\alpha^\lambda}\hat{\sigma}^\gamma\,,\\
\tensor{\left[a_1\right]}{_{\alpha}^{\beta^{\prime}}}={}&\left[\tensor{\hat{R}}{_\alpha^{\rho}}-\frac{1}{6}\hat{R}\delta_\alpha^\rho+\mu^2\tensor{\left(\tilde{g}^{-1}\right)}{_\alpha^{\rho}}\right]\tensor{\mathcal{P}}{_\rho^{\beta^\prime}}\,.\label{CovTayla1}
\end{align}
For the derivation of \eqref{CovTaylsig}--\eqref{CovTayla1}, we made use of \eqref{ColimSigT1}--\eqref{Colima1Ring}.
Inserting \eqref{CovTaylsig}--\eqref{CovTayla1} into \eqref{Factor1} and \eqref{Factor2} yields
\begin{align}
\Sigma^{\beta^\prime}_1(x^\prime,x)
={}&\mu\tensor{\mathcal{P}}{^{\beta^\prime}_\rho}\int_0^{\infty}\frac{\mathrm{d}s}{(4\pi s)^\omega}\hat{g}^{1/2}e^{-\frac{\hat{\sigma}}{2s}}\mathfrak{\hat{D}}^{1/2}\left\{\hat{\nabla}_{\alpha}\left(\tilde{g}^{-1}\right)^{\alpha\rho}-\frac{1}{2}\left(\tilde{g}^{-1}\right)^{\alpha}_{\nu}\tensor{\hat{R}}{_{\alpha\gamma}^{\nu\rho}}\hat{\sigma}^{\gamma}+\frac{1}{6}\left(\tilde{g}^{-1}\right)^{\alpha\rho}\hat{R}_{\alpha\gamma}\hat{\sigma}^{\gamma}\right.\nonumber\\
&\left.-\frac{1}{2s}\tensor{\left(\tilde{g}^{-1}\right)}{_\gamma^\rho}\hat{\sigma}^{\gamma}+\frac{1}{2}\left[\hat{R}^{\nu\rho}-\frac{1}{6}\hat{R}\hat{g}^{\nu\rho}+\mu^2\left(\tilde{g}^{-1}\right)^{\nu\rho}\right]\left(\tilde{g}^{-1}\right)_{\nu\gamma}\hat{\sigma}^{\gamma}\right\}+{\cal O}\left(\mathfrak{M}^4\right)\,,\label{Factor1Exp}\\\nonumber\\
\Sigma_{\beta^\prime}^2(x,x^\prime)={}&-\mu\int_0^{\infty}\frac{\mathrm{d}t}{(4\pi t)^\omega}\tilde{g}^{1/2}\mathfrak{\hat{D}}^{1/2}\left(\frac{\tilde{\mathfrak{D}}^{1/2}}{\mathfrak{\hat{D}}^{1/2}}e^{-\frac{\tilde{\sigma}}{2t}}\right)\left[\frac{\hat{\sigma}_\rho}{2t}+\frac{1}{4t}\tensor{\delta\Gamma}{_\rho_\gamma_\delta}\hat{\sigma}^\gamma\hat{\sigma}^\delta+\frac{1}{12t}\left(\tensor{\delta\Gamma}{_\rho_\gamma_\lambda}\tensor{\delta\Gamma}{^\lambda_\delta_\epsilon}-2\hat{\nabla}_\epsilon\tensor{\delta\Gamma}{_\rho_\gamma_\delta}\right)\hat{\sigma}^\gamma\hat{\sigma}^\delta\hat{\sigma}^\epsilon\right.\nonumber\\
&\left.-\frac{1}{6}\left(\tilde{g}^{-1}\right)^{\rho\gamma}\tilde{R}_{\gamma\delta}\hat{\sigma}^\delta+\frac{1}{12}\hat{\sigma}_{\rho}\tilde{R}\right]\tensor{\mathcal{P}}{^\rho_{\beta^\prime}}+{\cal O}\left(\mathfrak{M}^4\right)\label{Factor2Exp}\,,
\end{align}
where we have artificially separated a factor of $\mathfrak{\hat{D}}^{1/2}$ in \eqref{Factor2Exp}. 
This procedure of dealing with multiple propagators in a functional trace was proposed in \cite{Barvinsky1985}. Here, an additional complication is due to the presence of the two metrics $g_{\mu\nu}$ and $\tilde{g}_{\mu\nu}$, which in addition requires to expand the world function $\tilde{\sigma}$ in the exponent of the second product as well as the ratio $\tilde{\mathfrak{D}}^{1/2}/\mathfrak{\hat{D}}^{1/2}$,
\begin{align}
\frac{\tilde{\mathfrak{D}}^{1/2}}{\mathfrak{\hat{D}}^{1/2}}\exp\left(-\frac{\tilde{\sigma}}{2 t}\right)=\exp\left(-\frac{\tilde{g}_{\mu\nu}\hat{\sigma}^{\mu}\hat{\sigma}^{\nu}}{2 t}\right)&\left[1+\frac{1}{4t}\tilde{g}_{\alpha\lambda}\delta\Gamma^{\lambda}_{\beta\gamma}\hat{\sigma}^{\alpha}\hat{\sigma}^{\beta}\hat{\sigma}^{\gamma}-\frac{1}{48t}\left(4\tilde{g}_{\alpha\nu}\delta\Gamma^{\nu}_{\beta\lambda}\delta\Gamma^{\lambda}_{\gamma\delta}+3\tilde{g}_{\nu\lambda}\delta\Gamma^{\nu}_{\alpha\beta}\delta\Gamma^{\lambda}_{\gamma\delta}\right.\right.\nonumber\\
&\left.\left.+4\tilde{g}_{\alpha\lambda}\hat{\nabla}_{\beta}\delta\Gamma^{\lambda}_{\gamma\delta}\right)\hat{\sigma}^{\alpha}\hat{\sigma}^{\beta}\hat{\sigma}^{\gamma}\hat{\sigma}^{\delta}+\frac{1}{32 t^2}\tilde{g}_{\alpha\lambda}\tilde{g}_{\beta\nu}\delta\Gamma^{\lambda}_{\gamma\delta}\delta\Gamma^{\nu}_{\epsilon\eta}\hat{\sigma}^{\alpha}\hat{\sigma}^{\beta}\hat{\sigma}^{\gamma}\hat{\sigma}^{\delta}\hat{\sigma}^{\epsilon}\hat{\sigma}^{\eta}\right.\nonumber\\
&\left.+\frac{1}{12}\left(\tilde{R}_{\mu\nu}-\hat{R}_{\mu\nu}\right)\hat{\sigma}^{\mu}\hat{\sigma}^{\nu}\right]+\mathcal{O}\left(\mathfrak{M}^3\right)\,.\label{ExpVVExp}
\end{align} 
Finally, combining \eqref{Factor1Exp}, \eqref{Factor2Exp}, and the expansion \eqref{ExpVVExp}, we collect all terms up to order $\mathcal{O}\left(\mathfrak{M}^2\right)$ in a function $\Psi(s,t|\sigma^{\mu})$, which allows us to write the trace \eqref{T2convolution} as
\begin{align}
T_2=-\mu^2\int_{0}^{\infty}\frac{\mathrm{d}s\mathrm{d}t}{\left(4\pi\right)^{2\omega}s^\omega t^\omega}\int\mathrm{d}^{2\omega}x\,\mathrm{d}^{2\omega}x'\, \hat{\mathfrak{D}}\hat{g}^{1/2}\,\Psi(s,t|\hat{\sigma}^{\mu})\exp\left(-\frac{1}{4s}\tensor{G}{_\alpha_\beta}(s/t)\tensor{\hat{\sigma}}{^\alpha}\tensor{\hat{\sigma}}{^\beta}\right)\,.\label{T1Psi}
\end{align}
The main complexity related to the presence of the two metrics manifests in the ``interpolation metric'' $G_{\mu\nu}$ appearing in the exponential of \eqref{T1Psi}.
The interpolation metric relates the two metrics $\hat{g}_{\mu\nu}$ and $\tilde{g}_{\mu\nu}$ via the parameter $z$,
\begin{align}
\tensor{G}{_\mu_\nu}(z)\coloneqq\tensor{\hat{g}}{_\mu_\nu}+z\tensor{\tilde{g}}{_\mu_\nu}\,.
\end{align}
By construction,  $\Psi(s,t|\hat{\sigma}^{\mu})$ is a polynomial in $\hat{\sigma}^{\mu}(x,x')$,
\begin{align}
\Psi(s,t|\hat{\sigma}^{\mu})\coloneqq\hat{g}^{1/2}\sum_{k=2}^{8} s^{-k/2}\Psi^{(k)}_{\mu_1\dots\mu_k}(s,t)\hat{\sigma}^{\mu_1}\cdots\hat{\sigma}^{\mu_k}\,.\label{CovTaylorPsi}
\end{align}
The coefficients $\Psi^{(k)}_{\mu_1\dots\mu_k}(s,t)$ are local tensors parametrically depending on $s$ and $t$. The nonzero even coefficients are
\begin{align}
\Psi^{(2)}_{\alpha\beta}(s,t)={}&-\frac{1}{4t}\left(\tilde{g}^{-1}\right)_{\alpha\beta}-\frac{s}{4t}\left(\tilde{g}^{-1}\right)_{\lambda\eta}\tensor{\hat{R}}{^\lambda_\alpha^\eta_\beta}+\frac{s}{12t}\tensor{\left(\tilde{g}^{-2}\right)}{_{\alpha}^\eta}\tilde{R}_{\beta\eta}+\frac{1s}{3t}\left(\tilde{g}^{-1}\right)_{\alpha\lambda}\tensor{\hat{R}}{^\lambda_\beta}\nonumber\\
&-\frac{1}{24}\left( \tilde{g}^{-1}\right)_{\alpha\beta}\left(\frac{s}{t}\hat{R}+\tilde{R}\right)+\frac{s}{4t}\mu^2\left(\tilde{g}^{-2}\right)_{\alpha\beta}+\frac{s}{4t}\delta\Gamma^{\eta}_{\alpha\beta}\hat{\nabla}_{\lambda}\left(\tilde{g}^{-1}\right)^{\lambda}_{\eta}+\mathcal{O}\left(\mathfrak{M}^3\right),\\
\Psi^{(4)}_{\alpha\beta\gamma\delta}(s,t)={}&\frac{s}{8t}\tilde{g}_{\beta\eta}\delta\Gamma^{\eta}_{\gamma\delta}\hat{\nabla}_{\lambda}\left(\tilde{g}^{-1}\right)^{\lambda}_ {\alpha}-\frac{s}{24t}\left(\tilde{g}^{-1}\right)_{\alpha\lambda}\delta\Gamma^{\lambda}_{\beta\eta}\delta\Gamma^{\eta}_{\gamma\delta}+\frac{s}{12t}\left(\tilde{g}^{-1}\right)_{\alpha\lambda}\hat{\nabla}_{\beta}\delta\Gamma^{\lambda}_{\gamma\delta}\nonumber\\
&-\frac{s}{48\,t}\left(\tilde{g}^{-1}\right)_{\alpha\beta}\left(\tilde{R}_{\gamma\delta}-\hat{R}_{\gamma\delta}\right)+\mathcal{O}\left(\mathfrak{M}^3\right),\\
\Psi^{(6)}_{\alpha\beta\gamma\delta\mu\nu}(s,t)={}&\frac{s} {192t}\left(\tilde{g}^{-1}\right)_{\alpha\beta}\left(4\tilde{g}_{\gamma\lambda}\delta\Gamma^{\lambda}_{\delta\eta}\delta\Gamma^{\eta}_{\mu\nu}+3\tilde{g}_{\lambda\eta}\delta\Gamma^{\lambda}_{\gamma\delta}\delta\Gamma^{\eta}_{\mu\nu}+4\tilde{g}_{\gamma\lambda}\hat{\nabla}_{\delta}\delta\Gamma^{\lambda}_{\mu\nu}\right)\nonumber\\
&-\frac{s}{32t}\left(\tilde{g}^{-1}\right)_{\alpha\lambda}\tilde{g}_{\beta\eta}\delta\Gamma^{\lambda}_{\gamma\delta}\delta\Gamma^{\eta}_{\mu\nu}+\mathcal{O}\left(\mathfrak{M}^3\right),\\
\Psi^{(8)}_{\alpha\beta\gamma\delta\mu\nu\rho\sigma}(s,t)={}&-\frac{s}{128t}\left(\tilde{g}^{-1}\right)_{\alpha\beta}\tilde{g}_{\delta\lambda}\tilde{g}_{\gamma\eta}\delta\Gamma^{\lambda}_{\mu\nu}\delta\Gamma^{\eta}_{\rho\sigma}+\mathcal{O}\left(\mathfrak{M}^3\right)\,.
\end{align}
Changing integration variables from $x^{\mu'}\to \hat{\sigma}^\mu(x,x')$ leads to the Jacobian
\begin{align}
J={}&\det\left(\frac{\partial x^{\mu'}}{\partial\hat{\sigma}^{\nu}}\right)=\det\left(\hat{g}^{\rho\nu}\frac{\partial^2\hat{\sigma}(x,x')}{\partial x^{\mu'}\partial x^{\rho}}\right)^{-1}=\hat{g}^{1/2}(x)\hat{g}^{-1/2}(x')\mathfrak{\hat{D}}^{-1}(x,x')\,,
\end{align}
which cancels the factor $\mathfrak{\hat{D}}(x,x')\hat{g}^{1/2}(x')$ in \eqref{T1Psi}.
Thus, we write \eqref{T1Psi} as Gaussian integral over $\hat{\sigma}^{\mu}$,
\begin{align}
T_2=-\frac{\mu^2}{\left(4\pi\right)^{2\omega}}\int\mathrm{d}^{2\omega}x\hat{g}^{1/2}\int_{0}^{\infty}\frac{\mathrm{d}s\mathrm{d}t}{s^\omega t^\omega}\,\int\left(\prod_{\mu=1}^{2\omega}\mathrm{d}\hat{\sigma}^{\mu}\right)\, \Psi(s,t|\hat{\sigma}^\mu)\exp\left(-\frac{1}{4s}\tensor{G}{_\alpha_\beta}(s/t)\tensor{\hat{\sigma}}{^\alpha}\tensor{\hat{\sigma}}{^\beta}\right)\,.\label{T1Gauss}
\end{align}
We further perform a reparametrization $(s,t)\to(s,u)$, which allows to easily extract the divergent structure
\begin{align}
u =\frac{s}{t}\,,\quad  t=\frac{s}{u},\quad\det\left(\frac{\partial(s,t)}{\partial(s, u)}\right)=\frac{s}{u^2}\,.
\end{align}
All ultraviolet divergences are captured by the lower bound $s\to0$ of the $s$-integral and \eqref{T1Gauss} acquires the form
\begin{align}
T_2=-\frac{\mu^2}{\left(4\pi\right)^{2\omega}}\int\mathrm{d}^{2\omega}x\hat{g}^{1/2}\int_{0}^{\infty}\mathrm{d}u\,u^{\omega-2}\int_{0}^{\infty}\frac{\mathrm{d}s}{s^{2\omega-1}}\,\int\left(\prod_{\mu=1}^{2\omega}\mathrm{d}\hat{\sigma}^{\mu}\right)\, \Psi(u,s|\hat{\sigma}^\mu)\exp\left(-\frac{1}{4s}\tensor{G}{_\alpha_\beta}(u)\tensor{\hat{\sigma}}{^\alpha}\tensor{\hat{\sigma}}{^\beta}\right)\,.\label{T1GaussGu}
\end{align}
In order to extract the divergent part, we reparametrize the world function by absorbing a factor of $s^{-1/2}$,
\begin{align}
\hat{\sigma}\to\hat{\sigma}s^{1/2},\qquad\left(\prod_{\mu=1}^{2\omega}\mathrm{d}\hat{\sigma}^{\mu}\right)\to s^{\omega}\left(\prod_{\mu=1}^{2\omega}\mathrm{d}\hat{\sigma}^{\mu}\right)\,.
\end{align}
Finally, inserting the covariant Taylor expansion for $\Psi$, \eqref{T1GaussGu} reads
\begin{align}
T_2={}&-\frac{\mu^2}{\left(4\pi\right)^{2\omega}}\int\mathrm{d}^{2\omega}x\hat{g}^{1/2}\int_{0}^{\infty}\mathrm{d}u\,u^{\omega-2}\nonumber\\
\times&\sum_{k=0}^{4}\left[\int_{0}^{\infty}\frac{\mathrm{d}s}{s^{\omega-1}}\,\Psi^{(2k)}_{\mu_1\cdots\mu_{2k}}(u,s)\int\left(\prod_{\mu=1}^{2\omega}\mathrm{d}\hat{\sigma}^{\mu}\right)\, \hat{\sigma}^{u_1}\cdots\hat{\sigma}^{\mu_{2k}}\exp\left(-\frac{1}{4}\tensor{G}{_\alpha_\beta}(u)\tensor{\hat{\sigma}}{^\alpha}\tensor{\hat{\sigma}}{^\beta}\right)\right]\,.\label{T1DivIsolate}
\end{align}
Dimensional regularization annihilates all power law divergences and turns the logarithmically divergent $s$-integrals for $\omega\to2$ into poles $1/\varepsilon$ in dimension. Thus, we extract the divergent part of \eqref{T1DivIsolate} in $d=2\omega=4$ dimensions by collecting all terms in the integrand with total $s$-dependency $1/s$. For $\omega=2$, the prefactor is already of this form. Therefore, only the parts $\Psi^{(2k)}_{\mu_1...\mu_{2k}}(u)\coloneqq \Psi^{(2k)}_{\mu_1...\mu_{2k}}(0,u)$, which are independent of $s$, contribute to the divergent part
\begin{align}
T_2^{\mathrm{div}}={}&-\frac{\mu^2}{\left(4\pi\right)^{4}\varepsilon}\int\mathrm{d}^{4}x\,\hat{g}^{1/2}\int_{0}^{\infty}\mathrm{d}u\sum_{k=0}^{4}\left[\Psi^{(2k)}_{\mu_1...\mu_2k}(u)\int\left(\prod_{\mu=1}^{4}\mathrm{d}\hat{\sigma}^{\mu}\right)\, \hat{\sigma}^{u_1}\cdots\hat{\sigma}^{\mu_{2k}}\exp\left(-\frac{1}{4}\tensor{G}{_\alpha_\beta}(u)\tensor{\hat{\sigma}}{^\alpha}\tensor{\hat{\sigma}}{^\beta}\right)\right]\,.\label{T1DivPart}
\end{align}
Performing the Gaussian integrals yields
\begin{align}
&\frac{1}{\left(4\pi\right)^{2}}\int\left(\prod_{\mu=1}^{4}\mathrm{d}\hat{\sigma}^{\mu}\right)\,\hat{\sigma}^{\mu_1}\cdots\hat{\sigma}^{\mu_{2k}} \exp\left(-\frac{1}{4}\tensor{G}{_\alpha_\beta}\tensor{\hat{\sigma}}{^\alpha}\tensor{\hat{\sigma}}{^\beta}\right)=\frac{1}{G^{1/2}}\tensor*{\left[\operatorname{sym}_{k}\left(G^{-1}\right)\right]}{^{\mu_1\dots\mu_{2k}}}\,,
\end{align}
with the $k$th totally symmetrized power of the inverse interpolation metric $\left(G^{-1}\right)^{\mu\nu}$,
\begin{align}
\tensor*{\left[\operatorname{sym}_{k}\left(G^{-1}\right)\right]}{^{\mu_1\dots\mu_{2k}}}={}&\frac{(2k)!}{2^k k!}\left(G^{-1}\right)^{(\mu_1\mu_2}\cdots \left(G^{-1}\right)^{\mu_{2k-1}\mu_{2k})}\,.
\end{align}  
The fact that the Gaussian averages vanish for an odd number of $\hat{\sigma}^{\mu}$'s, a posteriori justifies that we have neglected these terms in the covariant Taylor expansion \eqref{CovTaylorPsi}.
Note that the resulting expressions are background tensors parametrically depending on $u$.
The divergent part of the trace \eqref{T1Trace} is given by
\begin{align}
T_2^{\mathrm{div}}={}&-\frac{\mu^2}{16\pi^2\varepsilon}\int\mathrm{d}^{4}x\,\hat{g}^{1/2}	\int_0^\infty\mathrm{d}u\sum_{k=0}^{4}\left[\frac{1}{G^{1/2}}\tensor*{\left[\operatorname{sym}_{k}\left(G^{-1}\right)\right]}{^{\mu_1\dots\mu_{2k}}}\Psi^{(2k)}_{\mu_1...\mu_{2k}}(u)\right]\,.
\end{align}	
Finally, the result for $T_{2}^{\mathrm{div}}$ can be expressed as linear combination
\begin{align}
T_2^{\mathrm{div}}={}&-\frac{\mu^2}{16\pi^2\varepsilon}\int\mathrm{d}^{4}x\hat{g}^{1/2}\sum_{k,\ell}C^{(2k,\ell)}_{\mu_1\dots\mu_{2k}}\,I^{\mu_1\dots\mu_{2k}}_{(2k,\ell)},\label{T2Result}
\end{align}
with  $u$-integrals 
\begin{align}
I^{\mu_1\dots\mu_{2k}}_{(2k,\ell)}\coloneqq\int_0^\infty\mathrm{ d}u \frac{\hat{g}^{1/2}}{G^{1/2}}u^\ell\tensor*{\left[\operatorname{sym}_{k}\left(G^{-1}\right)\right]}{^{\mu_1\dots\mu_{2k}}}\,,\label{IntegralsApp}
\end{align}
and $u$-independent coefficient tensors $C^{(2k,\ell)}_{\mu_1...\mu_{2k}}$,
\begin{align}
C^{(2,0)}_{\mu\nu}={}&-\frac{1}{12}\left(\tilde{g}^{-1}\right)_{\mu\nu}\tilde{R}+\frac{1}{6}\tensor{\left(\tilde{g}^{-2}\right)}{_{\mu}^{\beta}}\tilde{R}_{\nu\beta}\,,\\
C^{(2,1)}_{\mu\nu}={}&\frac{1}{2}\mu^2\tensor{\left(\tilde{g}^{-2}\right)}{_{\mu}_{\nu}}+\frac{2}{3}\left(\tilde{g}^{-1}\right)_{\mu}^{\alpha}\hat{R}_{\alpha\nu}-\frac{1}{12}\left(\tilde{g}^{-1}\right)_{\mu\nu}\hat{R}-\frac{1}{2}\left(\tilde{g}^{-1}\right)^{\alpha\beta}\hat{R}_{\mu\alpha\nu\beta}+\frac{1}{2}\delta\Gamma^{\alpha}_{\mu\nu}\hat{\nabla}_{\beta}\left(\tilde{g}^{-1}\right)_{\alpha}^{\beta}\,,\\
C^{(4,1)}_{\mu\nu\rho\sigma}={}&-\frac{1}{6}\delta\Gamma^{\alpha}_{\mu\nu}\delta\Gamma^{\beta}_{\rho\alpha}\left(\tilde{g}^{-1}\right)_{\sigma\beta}+\frac{1}{12}\left(\tilde{g}^{-1}\right)_{\mu\nu}\left(\hat{R}_{\rho\sigma}-\tilde{R}_{\rho\sigma}\right)+\frac{1}{3}\left(\tilde{g}^{-1}\right)_{\mu\alpha}\hat{\nabla}_{\rho}\delta\Gamma^{\alpha}_{\nu\sigma},\\
C^{(4,2)}_{\mu\nu\rho\sigma}={}&\frac{1}{2}\delta\Gamma^{\alpha}_{\mu\nu}\tilde{g}_{\alpha\sigma}\hat{\nabla}_{\beta}\left(\tilde{g}^{-1}\right)_{\rho}^{\beta}\,,\\
C^{(6,2)}_{\mu\nu\rho\sigma\alpha\beta}={}&\frac{1}{6}\delta\Gamma^{\lambda}_{\mu\nu}\delta\Gamma^{\eta}_{\sigma\lambda}\tilde{g}_{\rho\eta}\left(\tilde{g}^{-1}\right)_{\alpha\beta}+\frac{1}{8}\delta\Gamma^{\lambda}_{\mu\nu}\delta\Gamma^{\eta}_{\sigma\rho}\tilde{g}_{\lambda\eta}\left(\tilde{g}^{-1}\right)_{\alpha\beta}-\frac{1}{4}\delta\Gamma^{\lambda}_{\mu\nu}\delta\Gamma^{\eta}_{\sigma\rho}\tilde{g}_{\alpha\lambda}\left(\tilde{g}^{-1}\right)_{\beta\eta}\nonumber\\
&+\frac{1}{6}\tilde{g}_{\mu\lambda}\left(\tilde{g}^{-1}\right)_{\nu\sigma}\hat{\nabla}_{\beta}\delta\Gamma^{\lambda}_{\rho\alpha}\,,\\
C^{(8,3)}_{\mu\nu\rho\sigma\alpha\beta\gamma\delta}={}&-\frac{1}{8}\delta\Gamma^{\lambda}_{\mu\nu}\delta\Gamma^{\eta}_{\rho\sigma}\tilde{g}_{\alpha\lambda}\tilde{g}_{\beta\eta}\left(\tilde{g}^{-1}\right)_{\gamma\delta}\,.
\end{align}

\subsection{Divergent part of the fourth order trace}	
Since the trace \eqref{T2Trace} is $T_4={\cal O}(\mathfrak{M}^4)$, the operators in the trace $T_4$ can be freely commuted and we can use 
\begin{align}
\frac{\tensor*{\delta}{_\alpha^\beta}}{D_1}=\tensor*{\delta}{_\alpha^\beta}\frac{1}{\hat{\Delta}}+{\cal O}(\mathfrak{M})\,.
\end{align}
Explicitly, the divergent part of $T_4$ acquires the form
\begin{align}
T_4^{\mathrm{div}}={}&\mu^4\int{\rm d}^4x\left(\tilde{g}^{-2}\right)^{\alpha\beta}\left(\tilde{g}^{-2}\right)^{\gamma\delta}\hat{\nabla}_{\alpha}\hat{\nabla}_{\beta}\hat{\nabla}_{\gamma}\hat{\nabla}_{\delta}\frac{1}{\hat{\Delta}^2}\frac{1}{\tilde{\Delta}^2}\Big|_{x'=x}^{\mathrm{div}}\,.
\end{align} 
The trace $T_4^{\mathrm{div}}$ is evaluated in the same way as $T_2^{\mathrm{div}}$. We only state the final result
\begin{align}
T_4^{\mathrm{div}}=\frac{1}{16\pi^2\varepsilon}\int\mathrm{d}^4x\,\hat{g}^{1/2}\frac{\mu^4}{4}&\left[\tr\left(\tilde{g}^{-1}\right)\tr\left(\tilde{g}^{-2}\right)I_{(0,0)}-\tr\left(\tilde{g}^{-1}\right)\left(\tilde{g}^{-2}\right)^{\rho\sigma}\tensor{I}{_{(2,0)}_{\rho\sigma}}\right.\nonumber\\
&\left.-\tr\left(\tilde{g}^{-2}\right)\tensor{I}{_{(2,1)\alpha}^{\alpha}}+\left(\tilde{g}^{-2}\right)^{\mu\nu}\tensor{I}{_{(4,1)}_{\mu\nu}_{\alpha}^{\alpha}}\right]\,.\label{T4Result}
\end{align}

\section{FUNDAMENTAL INTEGRALS}\label{App:Integrals}
\subsection{Integral identities}
Starting from the fundamental integral identities
\begin{align}
\tensor*{I}{_{\ell}^\alpha^\beta^{\mu_1\cdots\mu_{2k}}}={}&-2\frac{\partial}{\partial\tensor{\tilde{g}}{_\alpha_\beta}}\tensor*{I}{_{\ell-1}^{\mu_1\cdots\mu_{2k}}}\nonumber\\
={}&-2\frac{\partial}{\partial\tensor{\hat{g}}{_\alpha_\beta}}\tensor*{I}{_\ell^{\mu_1\cdots\mu_{2k}}}\,,\label{FundInt}
\end{align}
it is possible to derive a sequence of useful integral identities
\begin{align}
\left(\hat{\nabla}_\lambda\tensor{\tilde{g}}{_\alpha_\beta}\right)\tensor*{I}{_{(2k+2,\ell)}^\alpha^\beta^{\mu_1\cdots\mu_{2k}}}={}&-2\hat{\nabla}_\lambda\tensor*{I}{_{(2k,\ell-1)}^{\mu_1\cdots\mu_{2k}}},\\
\tensor{\hat{g}}{_\alpha_\beta}\tensor*{I}{_{(2k+2,\ell)}^\alpha^\beta^{\mu_1\cdots\mu_{2k}}}={}&\begin{cases}
2(k-\ell)\,\tensor*{I}{_{(2k,\ell)}^{\mu_1\cdots\mu_{2k}}}&k>\ell\\[1mm]
2\tensor*{\left[\operatorname{sym}_{k} \left(\tilde{g}^{-1}\right)\right]}{^{\mu_1\dots\mu_{2k}}}& k=\ell
\end{cases}\,,\\
\left(\tilde{g}^{-1}\right)^{\nu}_{\alpha}\tensor*{I}{_{(2k+2,\ell)}^\alpha^{\mu_1\cdots\mu_{2k+1}}}={}&2k\left(\tilde{g}^{-1}\right)^{\nu(\mu_1}\tensor*{I}{_{(2k,\ell)}^{\mu_2\cdots\mu_{2k+1})}}-\tensor*{I}{_{(2k+2,\ell+1)}^{\nu\mu_1\cdots\mu_{2k+1}}}\,,\\
\tilde{g}^{\nu}_{\alpha}\tensor*{I}{_{(2k+2,\ell)}^\alpha^{\mu_1\cdots\mu_{2k+1}}}={}&2k\hat{g}^{\nu(\mu_1}\tensor*{I}{_{(2k,\ell-1)}^{\mu_2\cdots\mu_{2k+1})}}-\tensor*{I}{_{(2k+2,\ell-1)}^{\nu\mu_1\cdots\mu_{2k+1}}}\,,\\
\delta\Gamma_{\alpha\nu\beta}\tensor*{I}{_{(2k+2,\ell)}^{\alpha\beta}^{\mu_1\cdots\mu_{2k}}}={}&\hat{\nabla}_{\nu}\tensor*{I}{_{(2k,\ell)}^{\mu_1\cdots\mu_{2k}}}+2k\tensor{\delta\Gamma}{_{\lambda\nu}^{(\mu_1}}\tensor*{I}{_{(2k,\ell)}^{\mu_2\cdots\mu_{2k})\lambda}}\,.
\end{align}

\subsection{Evaluation of the integrals for the general case}
An interesting observation is that in $d=4$ dimensions the integrals \eqref{Integrals} can be evaluated in terms of invariants of the metric $\tilde{g}_{\mu\nu}$. The evaluation of all tensor integrals \eqref{Integrals} can be reduced to the evaluation of the fundamental scalar integral $I_{(0,0)}$.
In $d=4$ dimensions, the Cayley-Hamilton theorem guarantees that the eigenvalues $\lambda_{1},\dots,\lambda_4$ of $\tensor{\tilde{g}}{_{\mu}^{\nu}}$ can be expressed in terms of the invariants $e_k\coloneqq\tr\left(\tilde{g}^{-k}\right)$ with $k=1,\dots,4$. 
Therefore, the fundamental integral $I_{(0,0)}$ can be expressed in terms of the eigenvalues $\lambda_k(e_j)$,
\begin{align}
I_{(0,0)}=\int_0^\infty\mathrm{ d}u \frac{\hat{g}^{1/2}}{G^{1/2}(u)}=\int_0^\infty \frac{\mathrm{ d}u}{\sqrt{\prod_{k=1}^4\left(1+u\lambda_k\right)}}.\label{MotherIntegral}
\end{align}
The integral \eqref{MotherIntegral} can be evaluated explicitly and expressed in terms of the incomplete elliptic function of the first kind. The general integrals \eqref{Integrals} can then be obtained by differentiating the result with respect to $\tilde{g}_{\mu\nu}$ and $\hat{g}_{\mu\nu}$ and by making use of \eqref{FundInt}. We refrain from performing these operations, as the resulting expressions are horrendously complicated, impractical and not very illuminating. Instead, we choose to present the final result in terms of the much more compact integrals \eqref{Integrals}.

\subsection{Evaluation of the integrals for special cases}
In the case of the self-interacting vector field considered in Sec. \ref{SubSec:SelfIntVecField}, the integrals have the form
\begin{align}
I_{(2k,\ell)}^{\mu_1...\mu_{2k}}=\sum_{n=0}^{k}d_{(2k,\ell)}^{n}\hat{g}^{(\mu_1\mu_2}\cdots\hat{g}^{\mu_{2n-1}\mu_{2n}}\xi^{\mu_{2n+1}}\cdots\xi^{\mu_{2k})}\,.
\end{align}
The general coefficients $d_{(2k,\ell)}^{n}$ are given in a closed form in terms of the hypergeometric function ${}_2F_{1}$,
\begin{align}
d_{(2k,\ell)}^{n}={}&3^{-(\ell-1)/4}\frac{(2k)!}{2^n\,k!}\left(\begin{array}{c}k\\n\end{array}\right)\int_0^{\infty}\mathrm{d}u\,u^{\ell+k-n}(3+u)^{n-k-1/2}(1+u)^{-(k+3/2)}\nonumber\\
={}&2^{-n}\frac{(2k)!}{n!(k-n)!}3^{(\ell+1)/4}\left[\frac{\Gamma(k-\ell+1)\Gamma(\ell+1/2)}{\Gamma(k+3/2)} {}_{2}F_{1}(k-\ell+1,k-n+1/2,-\ell+1/2,3)+\right.\nonumber\\
&\left. 3^{\ell+1/2}\frac{\Gamma(-\ell-1/2)\Gamma(k+\ell-n+1)}{\Gamma(k-n+1/2)}\vphantom{\frac{\Gamma(k-\ell+1)\Gamma(l+1/2)}{\Gamma(k+3/2)}} {}_{2}F_{1}(k+3/2,k+\ell-n+1,\ell+3/2,3)\right]\,.
\end{align}
The hypergeometric function ${}_2F_{1}$ is defined as
\begin{align}
{}_2F_1(a,b,c,z) =  \sum_{k=0}^\infty\frac{\Gamma(c)}{ \Gamma(a) \,\Gamma(b)}\frac{\Gamma(a+k) \, \Gamma(b+k)}{\Gamma(c+k)} \frac{z^k}{k!}\,.
\end{align}
For the one-loop divergences, we only need the following coefficients 
\begin{alignat}{3}
&d_{(0,0)}^{0}={}\sqrt[4]{3}\left(-1+\sqrt{3}\right),\qquad
&&d_{(2,1)}^{0}={}\frac{22}{3}-4\sqrt{3}\,,
&&d_{(2,1)}^{1}={}-\frac{4}{3}+\sqrt{3}\,,\nonumber\\
&d_{(4,1)}^{0}={}\frac{4}{5}\left(73-42\sqrt{3}\right),
&&d_{(4,1)}^{1}={}\frac{4}{5}\left(-41+24\sqrt{3}\right),\qquad &&d_{(4,1)}^{2}={}\frac{1}{5}\left(7-3\sqrt{3}\right),\nonumber\\
&d_{(4,2)}^{0}={}\frac{4}{5}\sqrt[4]{3}\left(72-41\sqrt{3}\right),\qquad &&d_{(4,2)}^{1}={}\frac{8}{5}\sqrt[4]{3}\left(-27+16\sqrt{3}\right),\qquad &&d_{(4,2)}^{2}={}\frac{9}{5}\sqrt[4]{3}\left(2-\sqrt{3}\right).\label{d42}
\end{alignat}

\twocolumngrid

\bibliography{HKGenVecV2}{}

\end{document}